\documentclass[a4paper,10pt,nofootinbib]{revtex4}

\usepackage{algorithm}
\usepackage{algpseudocode}
\usepackage{amsfonts}
\usepackage{amsmath}
\usepackage{amssymb}
\usepackage{amsthm}
\usepackage{bm}
\usepackage{caption}
\usepackage{dcolumn}
\usepackage{graphicx,subcaption}
\usepackage{hyperref}

\newtheorem*{theorem*}{Theorem}
\newtheorem{theorem}{Theorem}
\newtheorem*{lemma*}{Lemma}
\newtheorem{lemma}{Lemma}

\begin{document}

\begin{abstract}
  Detector error models (DEMs) are commonly used to compile lower-level error models for simulating quantum error correction (QEC) syndromes (e.g. in the \texttt{stim} package); however, in recent years information has also begun to flow in the opposite direction, and DEMs estimated from syndromes are now assisting in understanding physical errors.
  We consolidate recent theoretical advances in DEM estimation and formalize several algorithms to learn DEM parameters and structure from syndromes without using a decoder, demonstrating recovery of known DEMs from simulated syndromes with precision limited only by finite-sample effects.
  We then applied these algorithms to estimate DEMs from Google's 72- and 105-qubit chips.
  Using a likelihood function that is tractable for small DEMs, we show that DEMs estimated directly from syndromes agree more closely with unseen syndromes than DEMs trained to optimize logical performance, whereas the latter outperform the former as priors for decoders in logical memory experiments.
  We used a time-series of estimated DEMs to track both global error and specific local errors over the course of a QEC experiment, suggesting applications in online characterization.
  We employ a sequence of DEM estimation techniques to discover and quantify long-range detector correlations spanning the width of the 105-qubit chip, for which DEM analysis suggests correlated measurement errors rather than high-weight Pauli errors as the most likely explanation.
  Finally, we present two artifacts in repetition code syndromes that are \emph{not} well-modeled by a DEM: correlated flipping of pairs of adjacent detectors in many consecutive rounds of QEC, and signatures consistent with radiation events occurring more frequently than previously reported.
  Although DEMs cannot capture all the relevant physics of a QEC device, we conclude that DEM estimation is poised to support hierarchical modeling by offering feedback to physical error models from syndromes.
\end{abstract}

\date{2026-03-11}
\title{Estimating Detector Error Models on Google's Willow}
\author{
  K.E. Arms,
  M.J. McHugh,\footnote{Corresponding Author: mjmchug@lps.umd.edu}
  J.E. Nyhan,
  W.F. Reus,
  J.L. Ulrich
}
\affiliation{Laboratory for Physical Sciences, 8050 Greenmead Dr, College Park, MD 20740, United States of America}

\maketitle

%%%%%%%%%%%%%%%%
% INTRODUCTION %
%%%%%%%%%%%%%%%%
\section{Introduction}
\label{sec:intro}
Over the last year, teams have published results directly demonstrating improved performance of logically encoded circuits \cite{reichardt2024demonstrationquantumcomputationerror} and increased lifetimes of error corrected states \cite{acharya2024quantumerrorcorrectionsurface}.
This leads naturally to an increased interest in the analysis of the resulting syndrome data and the description of the data generating process via detector error models (DEMs).

In the remainder of this paper, we discuss the development of DEMs as first class objects in QEC and their applications to QEC experiments.
We then introduce the mathematical notation that we use throughout the paper.
Subsequently, we define and analyze properties of DEMs.

We then provide algorithms for two learning tasks: a) rate estimation, in which the structure of the DEM is given and the goal is to learn the parameters from syndrome data-sets, and b) structure learning, where both structure and parameters must be learned.
For each task, we describe two algorithmic approaches, one based on moments of detectors and the other based on parities of the same.
In the last section, we demonstrate the application of these algorithms to recent Google experiments \cite{acharya2024quantumerrorcorrectionsurface} and on simulated data.
Specifically, we discuss the accuracy of decoding using estimated priors versus non-informative priors and priors trained to minimize the logical error rate (LER).
Then, we evaluate the agreement between various DEMs and syndromes using a method for estimating the likelihood of syndromes given a DEM, without requiring an underlying circuit noise model.
Next, we demonstrate the utility of DEM estimation in discovering and investigating errors not explained by typical circuit noise.
Finally, we depart from the direct discussion of DEMs to present large, anomalous correlations in the Google data.

%% Subsection: Prior Work
\subsection{Prior Work}
\label{subsec:prior-work}
We now present a brief, incomplete chronology of the concept of detector error models and error estimation using syndrome data.
We formally define detectors \cite{2024DerksDesigning} in the context of QEC below, but informally, a detector is a parity constraint among measurement outcomes that is deterministic in the absence of noise.
A given class of (detectable) errors violates, or flips, at least one parity constraint, and a DEM specifies a)~the set of detectors flipped by each error class---the \emph{structure}---and b)~the independent rate of each error class---the \emph{parameters}.

The DEM estimation problem was first directly addressed for a DEM with graphical structure (in which each error class flips at most a pair of detectors) in \cite{https://doi.org/10.1002/qute.201800012}, constructing the analytically estimable $p_{ij}$ which appear in later works.
This work is extended in \cite{PhysRevLett.128.110504}, which includes a concise description of the DEM in terms of a decoding \emph{hyper}graph in Section II and describes an algorithm for estimation of error parameters of a fixed hypergraph in Supplement G.
These approaches to the problem of rate estimation are the conceptual ancestors of one of the two classes of algorithms discussed here: those based on moments of detectors.

Meanwhile, the DEM was identified as a first class object for stabilizer simulation and QEC decoding via minimum-weight perfect matching in \cite{gidney2021stim} and \cite{Higgot_2025} respectively.
We note that the definitions adopted in these works are the ones most closely followed in our work due primarily to use of the tool chain described therein.

In \cite{PhysRevResearch.3.013292}, the authors discuss many of the ideas covered here and present a Pauli error-process estimator utilizing a modified belief-propagation (BP) decoder and expectation-maximization (EM) algorithm.
However, in that and a following work \cite{Wagner2022paulichannelscanbe}, they limit their approach to the ``framework of quantum-data syndrome codes'' where Pauli errors may occur only on data qubits and syndrome measurements may err.
This restriction is not observed in our work.

In the past year, there have been several machine-learning based, decoder-in-the-loop algorithms designed to optimize the syndrome-extraction DEM for logical memory performance.
The algorithms we discuss here do not rely on decoders and offer more direct insight into the syndrome statistics of QEC devices, at the expense of logical fidelity.

Two recent works mark an important advance in understanding DEMs: the analysis of detector parities, rather than moments, for DEM learning.
Remm \emph{et al.} \cite{remm2025} introduced a parity-based algorithm for rate estimation and applied it to estimate a DEM from syndromes generated by a laboratory QEC chip.
By estimating rates for a DEM with hand-picked structure, the authors diagnose leakage and multi-qubit errors in the chip.
The recent work of \cite{blumekohout2025estimatingdetectorerrormodels}, independent of \cite{remm2025}, made great theoretical strides in relating parities of sets of detectors to DEM parameters via a Walsh-Hadamard transform, thereby linking DEM research to a rich body of literature and laying the groundwork for general algorithms for both rate estimation and structure learning.

%% Subsection: Notation, Typography and Bijections
\subsection{Notation, Typography and Bijections}
\label{subsec:notation}
In the field of DEM analysis, we have found it prudent to adopt concise notation. In order to make the following analyses easy to follow, we define the following notations and attempt consistency throughout.
\begin{itemize}
\item $[\text{a logical statement}]$: the \textit{indicator function} for the logical statement being true. Also known as \textit{Iverson brackets}.
\item $[n] \equiv \{0, 1, 2, ..., n-1\}$: the $n$-set. The set of all non-negative integers below $n$.
\item $\mathcal{P}{S}$: the \textit{power-set}, the set of all subsets of $S$.
\end{itemize}

It is often useful to represent an object in three different ways: as a integer, as a bit-vector (little-endian) representing that integer, and as a set, whose elements represent ``which bits in the bit-vector are 1s.'' This leads us to define the following mappings between these representations.
\begin{itemize}
\item $f: \mathbb{F}_{2^n} \to \mathbb{F}_{2}^{n}$: the bijection between a non-negative integer below $2^n$ and bit-vectors of length $n$ described by little-endian binary representation.
\item $g: \mathbb{F}_2^n \to \mathcal{P}{[n]}$: the bijection from bit-vectors $\mathbf{x}$ of length $n$ to $S \subseteq [n]$ where bit $x_i = [i \in S]$.
\item $h: \mathbb{F}_{2^n} \to \mathcal{P}{[n]}$: the composition of the two above functions, i.e. $h(\cdot) = g(f(\cdot))$.
\end{itemize}

Given the above, we use typesetting to differentiate between different quantities related by these functions:
\begin{itemize}
\item $a \in \mathbb{F}_{2^n}$: the integer representation; lowercase, italicized.
\item $\mathbf{a} = f(a)$: the little-endian bit-vector representation; lowercase, boldface.
\item $A = h(a) = g(\mathbf{a})$: the set of locations of 1s in the bit-vector; uppercase, italicized.
\end{itemize}

With these definitions, we can articulate the following identities:
\begin{itemize}
\item The bit-wise product:
  \begin{align}
    \mathbf{a \cdot b} \equiv \bigoplus_{i \in [n]} a_i b_i = |A \cap B| \bmod 2
  \end{align}
\item ``Element-wise less than or equal'': for all $i$, is $a_i$ less than $b_i$?
  \begin{align}
    \mathbf{a \le b} \equiv \prod_{i \in [n]} [a_i \le b_i] = [A \subseteq B]
  \end{align}
  Additionally, the latter trivially implies
  \begin{align}
    A \subseteq B \implies a \leq b, \label{eq:linear-extension}
  \end{align}
  indicating that little-endian ordering is a linear extension of the poset of power-sets under inclusion.
\end{itemize}

We have the following shorthand based on the above bijections:
\begin{itemize}
\item $\mathbf{a} \in \mathbf{X}$ for $\mathbf{a} \in \mathbb{F}_2^n, \mathbf{X} \in \mathbb{F}_2^{n \times m}$ means that $\mathbf{a}$ is a column of $\mathbf{X}$.
\item $a \in \mathbf{X}$ for $a \in \mathbb{F}_{2^n}, \mathbf{X} \in \mathbb{F}_2^{n \times m}$ means that $\mathbf{a} = f(a)$ is a column of $\mathbf{X}$.
\item $A \in \mathbf{X}$ for $A \in \mathcal{P}{[n]}, \mathbf{X} \in \mathbb{F}_2^{n \times m}$ means that $\mathbf{a} = g^{-1}(A)$ is a column of $\mathbf{X}$.
\end{itemize}

We define the Hadamard matrix, $\mathbf{H}$, recursively, to obey $\mathbf{HH} = 2^{n}\mathbf{I}$ which yields
\begin{align*}
    \mathbf{H}^{(0)} &= (1), \\
    \mathbf{H}^{(n + 1)}
    &= \begin{pmatrix}
        \mathbf{H}^{(n)} & \mathbf{H}^{(n)} \\
        \mathbf{H}^{(n)} & -\mathbf{H}^{(n)} \\
    \end{pmatrix}.
\end{align*}

%%%%%%%%%%%%%%%%%%%%%%%%%%%%%
% DETECTOR ERROR MODEL MATH %
%%%%%%%%%%%%%%%%%%%%%%%%%%%%%
\section{Detector Error Models and some Mathematical Properties}
\label{sec:dems}
We are indebted to \cite{2024DerksDesigning, blumekohout2025estimatingdetectorerrormodels, gidney2021stim, Higgot_2025} from which we liberally borrow.
We remind the reader that a detector is a parity constraint on a series of measurement outcomes in a QEC circuit.
For the repetition and surface codes discussed here, a detector is typically the mod-2 sum of the same stabilizer measurement in successive rounds of syndrome extraction.
Thus, in the time-bulk of a memory experiment (i.e.~except for the initial and final rounds), a detector exists for each stabilizer in each round and takes the value one if and only if the corresponding stabilizer measurement changed in that round.
We dispense with raw stabilizer measurements and define a \emph{syndrome} as the sequence of detector values for one replicate, or \emph{shot}, of a multi-round, logical memory experiment.

The \emph{detector error model} (DEM) describes a data generating process for a syndrome, $\mathbf{x} \in \mathbb{F}_2^n$.
The DEM is defined as a binary \emph{incidence matrix}, $\mathbf{M} \in \mathbb{F}_2^{n \times E}$ and a real \emph{excitation rate vector}, $\boldsymbol{\theta} \in [0, 1]^E$.
Column $\mathbf{s}_i$ of this matrix represents a \emph{hyperedge} linking a potentially arbitrary subset of detectors flipped by the $i$-th excitation.
The only constraint we place on a DEM is that the columns of $\mathbf{M}$ be unique.
In general $1 \le E \le 2^n - 1$, but typically $E$ is polynomial in the number of gates performed in the QEC circuit.

Given $(\mathbf{M}, \boldsymbol{\theta})$, we generate data by drawing a random \emph{excitation vector} $\mathbf{e}$ with $e_i \sim$ Bernoulli($\theta_i$).
Then $\mathbf{x} = \mathbf{Me},$ where the matrix-vector multiply is performed in $\mathbb{F}_2$.
This is essentially the mathematical formulation for the process of QEC simulation in \texttt{stim} \cite{gidney2021stim}.
Linear algebra in $\mathbb{F}_2$ enforces the English language statement, ``the probability a given detector flips is the probability an odd number of excited hyperedges touch a detector.''
Importantly, in this formulation the elements of $\mathbf{e}$ are all \emph{independent}.

We discuss two classes of algorithms for estimating $\mathbf{M}$ and $\boldsymbol{\theta}$ from syndrome data.
\emph{Rate estimation} is any algorithm by which one determines the values of $\mathbf{\theta}$ for a DEM with a specifed $\mathbf{M}$.
In other cases, we imagine that $\mathbf{M}$ is the $n$ bit Hamming matrix representing all possible hyperedges. Now we want to find \emph{which} $\theta_i$ are non-zero \emph{and} estimate these rates.
We call these algorithms \emph{structure learning}.
For both rate estimation and structure learning, we consider algorithms based on moments and parities of the data-set.
We show that both approaches agree on simulated data.

%% Subsection: Moments, Parities, Polarizaitons and More
\subsection{Moments, Parities, Polarizations and More}
\label{subsec:moments-and-more}
As much of this paper discusses estimating DEM properties from syndromes (i.e.~random bit-vectors), it is worthwhile to discuss some generic statistics of these random bit-vectors.
We consider two data-generating models; the first is fully general, $\mathbf x \sim \text{Categorical}(\mathbf{p})$ subject to $\sum_x p_x = 1$ and $0 \le p_x \le 1$ for all $x$.
The second is the detector error model.
For a given DEM the probability of a syndrome $\mathbf{x}$ is given by
\begin{align}
  p_\mathbf{x} 
  = \sum_{\mathbf{e}\in\mathbb{F}_2^E} [\mathbf{x} = \mathbf{M} \mathbf{e}]\Pr(\mathbf{e} | \boldsymbol{\theta}) 
  = \sum_{\mathbf{e}\in\mathbb{F}_2^E} [\mathbf{x} = \mathbf{M} \mathbf{e}]\prod_{i\in[E]}\theta_i^{e_i}(1 - \theta_i)^{1 \oplus e_i}.
\end{align}
The final equality is due to the assumed independence of the excitation elements.

A related quantity is the \emph{moment} of $\mathbf{y}$,\footnote{Throughout this paper, we use $\mathbf y$ ($Y$) to denote a bit-vector (set of detectors) associated with some observable quantity, whereas $\mathbf s$ ($S$) represents a bit-vector (set of detectors) corresponding to a DEM hyperedge. At times, it is necessary to introduce other symbols to fill roles in these categories, but we endeavor to preserve this convention where possible.} defined as the probability that all bits indicated by
$\mathbf{y}$ are one;
\begin{align}
  \mu_\mathbf{y} \equiv \langle [\mathbf{y} \le \mathbf{x}] \rangle
  = \sum_{\mathbf{x} \in \mathbb{F}_2^n} [\mathbf{y} \le \mathbf{x}] p_\mathbf{x}
  = \sum_{\mathbf{e}\in\mathbb{F}_2^E} [\mathbf{y} \le \mathbf{M} \mathbf{e}] \Pr(\mathbf{e} | \boldsymbol{\theta}) 
  = \sum_{\mathbf{e}\in\mathbb{F}_2^E} [\mathbf{y} \le \mathbf{M} \mathbf{e}]
  \prod_{i\in[E]} \theta_i^{e_i}(1 - \theta_i)^{1 \oplus e_i}.
\end{align}
We note that many columns in $\mathbf{M}$ do not intersect with $\mathbf{y}$ and thus can be averaged over without impacting the moment of interest.
We define the neighborhood $\mathcal{N}(Y)$ as the set of hyperedges (columns of the incidence matrix, $\mathbf{M}$) which overlap with $Y$.
We then define $\mathbf{M}^{\prime} \in \mathbb{F}_2^{|Y|\times|\mathcal{N}(Y)|}$ to be the sub-matrix of the incidence matrix formed by restricting $\mathbf{M}$ to the rows indicated by $\mathbf{y}$ and the columns which have any ones in those rows.
Correspondingly, $\mathbf{e}^{\prime} \in \mathbb{F}_2^{|\mathcal{N}(Y)|}$ is an excitation vector defined only over those rows.
Then we may write
\begin{align}
  \mu_Y
  &= \sum_{\mathbf{e}^{\prime} \in \mathbb{F}_2^{|\mathcal{N}(Y)|}}
  [\mathbf{1} = \mathbf{M}^{\prime}\mathbf{e}^{\prime}]
  \prod_{i \in \mathcal{N}(Y)} \theta_i^{e_i}(1 - \theta_i)^{1 \oplus e_i}. \label{eq:neighborhood-moment}
\end{align}
This is equivalent to Equation 5 in \cite{PhysRevLett.128.110504} from which they derive their estimation algorithm.
These equations also form the basis for the recent work \cite{takou2025estimatingdecodinggraphshypergraphs}, however the authors do not appear to have directly addressed (nor significantly suffered from) the issue of potentially many excitations now having the same signature within a moment.
We derive a distinct estimation algorithm from both approaches using \autoref{eq:neighborhood-moment}.

Following \cite{blumekohout2025estimatingdetectorerrormodels} we examine the parity of a subset of detectors indicated by $\mathbf{y}$, but under the group isomorphism $\{\{0, 1\}, \oplus\} \to \{\{+1, -1\}, \times\}$,
\begin{align}
  z_\mathbf{y}(\mathbf x)\equiv(-1)^{\mathbf{x\cdot y}} = 1-2\mathbf{x\cdot y}. \label{eq:polarization-definition}
\end{align}
We define the polarization to be the expectation of the parity,
\begin{align}
  \pi_\mathbf{y}\equiv\langle z_\mathbf{y}\rangle_\mathbf{x}=\sum_{\mathbf{x}\in\mathbb{F}_2^n}(-1)^{\mathbf{x\cdot y}}p_\mathbf{x}.
\end{align}
This implies a Hadamard transform relates polarizations and probabilities
\begin{align}
  \boldsymbol{\pi} = \mathbf{Hp}. \label{eq:parity-is-hadamard-probs}
\end{align}
Applying the inverse, we have $2^{-n}\mathbf{H}\boldsymbol{\pi} = \mathbf{p}$.
Now consider the parity under a DEM
\begin{align}
  z_\mathbf{y}(\mathbf{x}) = 1 - 2 \mathbf{y \cdot x} = 1 - 2 \mathbf{y}^T\mathbf{Me} = \prod_{\mathbf{s} \in \mathbf{M}} (1 - 2 e_\mathbf{s} \mathbf{s \cdot y}).
\end{align}
Since excitations are independent, we see that the polarization is:
\begin{align}
  \pi_{\mathbf y}=\prod_{\mathbf{s}\in\mathbf{M}}(1-2\theta_\mathbf{s})^{\mathbf{s\cdot y}}. \label{eq:dem-polarizations}
\end{align}

As in \cite{blumekohout2025estimatingdetectorerrormodels}, we define the depolarization,
\begin{align}
  \omega_{\mathbf y} \equiv -\ln \pi_{\mathbf y}, \label{eq:polarization-to-depolarization}
\end{align}
and the attenuation,
\begin{align}
  \psi_{\mathbf s} = - \ln (1 - 2\theta_{\mathbf s}). \label{eq:rate-to-attenuation}
\end{align}
Applying the negative logarithm to both sides of \autoref{eq:dem-polarizations} yields
\begin{align}
  \omega_\mathbf{y} &= \sum_{\mathbf{s} \in \mathbb{F}_2^n} \mathbf{y\cdot s}\,\psi_\mathbf{s}, \\
  \omega_Y &= \sum_{\emptyset \subseteq S \subseteq [n]} (|Y \cap S| \bmod 2) \psi_S. \label{eq:theorem-1-hypothesis-sums} \\
  \boldsymbol{\omega} &= \mathbf{W}\boldsymbol{\psi}. \label{eq:omega-is-Wpsi}
\end{align}
The last equation is just the linear-algebraic restatement of the previous two.
This leads to a useful theorem.
\begin{theorem}
  \label{theorem:GZpsi-is-Lomega}
  If \autoref{eq:theorem-1-hypothesis-sums} is true for all $A \subseteq [n]$, then for any $S \subseteq [n]$
  \begin{align}
    \label{eq:GZpsi-is-Lomega_sums}
    \frac{2^{|S|}}{-2}\sum_{S\subseteq A\subseteq[n]}\psi_A=\sum_{\emptyset\subseteq B\subseteq S}(-1)^{|B|}\omega_B.
  \end{align}
\end{theorem}
We give the proof in Appendix \ref{sec:proofs}.

This theorem has two interesting consequences.
First, it provides a derivation of Equation 10 of \cite{remm2025}, which the authors verify but do not derive.
Exponentiation of \autoref{eq:GZpsi-is-Lomega_sums} yields,
\begin{align}
  \left(\prod_{S \subseteq A \subseteq [n]} (1 - 2\theta_A)\right)^{(-1)2^{|S| - 1}}
  &= \prod_{\emptyset \subseteq B \subseteq S} \pi_B^{(-1)^{|B|}} \\
  \prod_{S \subseteq A \subseteq [n]} (1 - 2\theta_A)
  &= \prod_{\emptyset \subseteq B \subseteq S} \pi_B^{(-1)^{|B| - 1} 2^{1 - |S|}} \\
  (1 - 2\theta_S)
  &=
  \frac{\prod_{\emptyset \subseteq B \subseteq S} \pi_B^{(-1)^{|B| - 1} 2^{1 - |S|}}}
       {\prod_{S \subset A \subseteq [n]} (1 - 2\theta_A)} \\
       \theta_S
       &= \frac{1}{2} - \frac{1}{2}
       \frac{\prod_{\emptyset \subseteq B \subseteq S} \pi_B^{(-1)^{|B| - 1} 2^{1 - |S|}}}
            {\prod_{S \subset A \subseteq [n]} (1 - 2\theta_A)}.
            \label{eq:remm-equation-10}
\end{align}
Next, we write the $2^n$ equations implied by \autoref{eq:GZpsi-is-Lomega_sums} as a matrix equation:
\begin{align}
  \label{eq:GZpsi-is-Lomega}
  \mathbf{GZ}\boldsymbol{\psi} = \mathbf{L} \boldsymbol{\omega}.
\end{align}
We discuss these $2^n \times 2^n$ matrices in Table \ref{tab:matrix-properties}.
We note that $\mathbf{G}$ is a diagonal matrix of the first $2^n$ terms of Gould's sequence (scaled by -1/2).
$\mathbf{Z}$ is a matrix representation of the $\zeta$-function on the Boolean poset, as such, it is upper triangular and invertible.
We prove the recursion for $\mathbf{L}$ in Appendix \ref{sec:proofs}.
A similar proof (not shown) serves for $\mathbf{G}$ and $\mathbf{Z}$.
Moreover, we demonstrate that $\mathbf{LL} = \mathbf{I}$, leading us to rewrite \autoref{eq:GZpsi-is-Lomega} as
\begin{align}
  \mathbf{LGZ}\boldsymbol{\psi} = \boldsymbol{\omega}.
\end{align}
We also show in Appendix \ref{sec:proofs}, that $\mathbf{LGZ} = -\mathbf{H} / 2$.
Thus,
\begin{align}
  \frac{-1}{2}\mathbf{H}\boldsymbol{\psi}
  &= \boldsymbol{\omega}. \label{eq:Hpsi-is-omega} \\
  \therefore\;\; \boldsymbol{\psi}
  &= \frac{-2}{2^n} \mathbf{H}\boldsymbol{\omega}. \label{eq:psi-is-Homega}
\end{align}
\autoref{eq:psi-is-Homega} is equivalent to Equation 31 in \cite{blumekohout2025estimatingdetectorerrormodels} accounting for a different normalization convention and variable names.

We note that the LHS of \autoref{eq:Hpsi-is-omega} and RHS of \autoref{eq:omega-is-Wpsi} are not equal for arbitrary $\boldsymbol{\psi}$.
The authors of \cite{blumekohout2025estimatingdetectorerrormodels} note that $\mathbf W$ is not invertible, as it annihilates the vector $(1,0,0,\cdots, 0)$.
This leaves $\psi_0$ as a free parameter.
Applying \autoref{eq:GZpsi-is-Lomega_sums} with $Y = \emptyset$ yields
\begin{align}
  0 = \omega_\emptyset = \frac{1}{-2} \sum_{\emptyset \subseteq S \subseteq [n]} \psi_S \implies \psi_\emptyset = - \sum_{\emptyset \subset S \subseteq [n]} \psi_S.
\end{align}
Using this value of $\psi_0$, \autoref{eq:Hpsi-is-omega} and \autoref{eq:omega-is-Wpsi} yield equivalent $\boldsymbol{\omega}$.
The authors of \cite{blumekohout2025estimatingdetectorerrormodels} define the negative of this quantity as the \emph{total attenuation}.
By taking this value for an otherwise unused parameter, we exchange the non-invertible binary Walsh transform for the traditional Hadamard transform and obtain a bijection from attenuations to depolarizations.

\begin{table}
  \centering
  \begin{tabular}{c|ccccc}
    $\mathbf{X}$ & $\mathbf{G}$ & $\mathbf{Z}$ & $\mathbf{L}$ & $\mathbf{H}$ & $\mathbf{W}$ \\
    \hline
    $X_{ij}$
    & $- \frac{2^{|I|}}{2} \delta_{ij}$
    & $[I \subseteq J]$
    & $(-1)^{|J|} [J \subseteq I]$
    & $(-1)^{|I \cap J|}$
    & $|I \cap J| \% 2$ \\
    $\mathbf{X}^{(0)}$ & (-1/2) & (1) & (1) & (1) & (0) \\
    $\mathbf{X}^{(n+1)}$
    & $\begin{pmatrix} \mathbf{G}^{(n)} & \mathbf{0} \\ \mathbf{0} & 2 \mathbf{G}^{(n)} \end{pmatrix}$
    & $\begin{pmatrix} \mathbf{Z}^{(n)} & \mathbf{Z}^{(n)} \\ \mathbf{0} & \mathbf{Z}^{(n)} \end{pmatrix}$
    & $\begin{pmatrix} \mathbf{L}^{(n)} & \mathbf{0} \\ \mathbf{L}^{(n)} & -\mathbf{L}^{(n)} \end{pmatrix}$
    & $\begin{pmatrix} \mathbf{H}^{(n)} & \mathbf{H}^{(n)} \\ \mathbf{H}^{(n)} & -\mathbf{H}^{(n)} \end{pmatrix}$
    & $\begin{pmatrix} \mathbf{W}^{(n)} & \mathbf{W}^{(n)} \\ \mathbf{W}^{(n)} & 1 \oplus \mathbf{W}^{(n)} \end{pmatrix}$ \\
  \end{tabular}
  \caption{A description of the matrices which appear frequently in this work.
    The table indicates their entries for arbitrary $i, j \in Z_{2^n}$ and their recursions starting from the case $n = 0$.
    The recursions are derivable from the definition of the entries. Note that $I = h(i)$, $J =h(j)$.
  }
  \label{tab:matrix-properties}
\end{table}

%% Subsection: Analytic Syndrome Likelihoods
\subsection{Analytic Syndrome Likelihoods}
\label{subsec:likelihood}
We now discuss the use of the DEM formalism to construct likelihood functions for syndrome data-sets.
We also suggest their applicability for hypothesis testing and goodness-of-fit metrics.

Suppose we have a probability vector $\mathbf{p}$.
We may combine Equations \ref{eq:psi-is-Homega}, \ref{eq:parity-is-hadamard-probs}, \ref{eq:rate-to-attenuation} and \ref{eq:polarization-to-depolarization} to construct an excitation rate vector $\boldsymbol{\theta}$ analytically with the transform,
\begin{align}
  \boldsymbol{\theta}
  &= \frac{1}{2} - \frac{1}{2} \exp\left\{
  \frac{-2}{2^n} \mathbf{H}\left[ -\ln \left( \mathbf{H[p]} \right) \right] \right\}.
  \label{eq:excitation-rates-from-probability}
\end{align}
This is a real value when all entries of $\boldsymbol{\pi} = \mathbf{H[p]}$ are positive.
Similarly, if provided an excitation vector $\boldsymbol{\theta}$ (with appropriate value for $\theta_0$), we may recover the probability vector,
\begin{align}
  \mathbf{p} = \frac{1}{2^n} \mathbf{H}\left[
    \exp \left\{
    \frac{1}{2} \mathbf{H}\left[ - \ln\left(1 - 2 \boldsymbol{\theta}\right)\right]
    \right\}
    \right].
  \label{eq:probability-from-excitation-rates}
\end{align}
The only requirement here is that $0 \le \theta_i < 1/2 $ for all $i > 0$. In both equations, the natural logarithm and exponentiation are applied element-wise to the vector arguments.

\autoref{eq:probability-from-excitation-rates} in the context of QEC is the likelihood of a syndrome assuming the generating DEM.
When $n$ is not too large, we can calculate the likelihood exactly and compare it to the empirical distribution for a measure of model-agreement.
However, even for larger syndromes where the exponential cost of the outer Hadamard transform is prohibitive, one can always look at a reduced dimensional marginal of calculable size and determine:
\begin{align}
  \Pr(\mathbf{x}_S = \mathbf{a})
  &= \frac{1}{2^{|S|}} \sum_{\mathbf{y} \in \mathbb{F}_2^{|S|}} (-1)^{\mathbf{a\cdot y}}\exp\left[- \sum_{\mathbf{b}\in \mathbb{F}_2^{|S|}} \mathbf{y\cdot b} \psi_\mathbf{b}^\star \right],
\end{align}
where we define $\mathbf{x}_S = [x_{i_1}, x_{i_2}, ... x_{i_{|S|}}]$, the bit vector constructed by only looking at the bits with indices in $S$, and
\begin{align}
  \psi_\mathbf{b}^\star
  = \sum_{\mathbf{c} \in \mathbf{Z}_2^n} [\mathbf{c}_S = \mathbf{b}] \psi_\mathbf{c},
\end{align}
the summed attenuation for all excitations which have signature $\mathbf{b}$ when only looking at bits in $S$.

Choosing different sets of bits for which to estimate marginal likelihoods and calculating the likelihood from the DEM may be a good way of testing model violation.
In the future, we hope to examine whether it is possible to combine these marginal likelihoods into a global score of model agreement.

%% Subsection: DEM Constraints
\subsection{DEM Constraints}
\label{subsec:dem-constraints}
We now discuss some observations about the class of probability distributions represented by physical DEMs and both a successful and unsuccessful application of the formalism.
A DEM is \emph{physical} when all excitation rates are small (below 1/2).
We have yet to determine the bounds of probability distributions represented by physical DEMs and would welcome a discussion thereof.

We have found one important constraint imposed by physical DEMs; they cannot predict pairwise anti-correlations.
To see this, assume one has only two detectors and excitation rates $0 \le \theta_{10}, \theta_{01}, \theta_{11} \le 1 / 2$.
The covariance between the two bits is defined as
\begin{align*}
  \mathsf{Cov}
  \equiv \mu_{\{0, 1\}} - \mu_{\{0\}} \mu_{\{1\}}
  = \theta_{11}(1 - \theta_{11})\left[1 - 2(\theta_{01} + \theta_{10}) + 4 \theta_{01}\theta_{10} \right].
\end{align*}
One can easily see $0 \le \theta_{11}(1 - \theta_{11}) \le 1/4$ from simple maximization.
Applying the extreme value theorem to the function
\begin{align*}
  f(\theta_{01}, \theta_{10}) = 1 - 2(\theta_{01} + \theta_{10}) + 4 \theta_{01}\theta_{10}
\end{align*}
shows $0 \le f(\theta_{01}, \theta_{10}) \le 1$. Thus $0 \le \mathsf{Cov} \le 1/4$.
Therefore, in a two-bit DEM, small errors cannot produce anti-correlations between the two detectors.
In Section 3.1 of \cite{blumekohout2025estimatingdetectorerrormodels}, the authors show that the covariance between detectors $i$ and $j$ is proportional to the \emph{aggregated attenuation} of event $\lbrace i, j \rbrace$, which is in turn the sum of attenuations for all DEM events for which $i, j = 1$;
\begin{align}
  \mathsf{Cov}(i, j) &\propto \psi_{\lbrace i, j \rbrace}^\star = \sum_{\mathbf s \colon s_i, s_j = 1} \psi_{\mathbf s}.
\end{align}
Recall that $ 0 \le \theta_{\mathbf s} \le 1/2 \iff 0 \le \psi_{\mathbf s}$; that is, physical DEMs have non-negative, real attenuations.
Since attenuations can only combine additively to produce pairwise, aggregated attenuations, physical DEMs have no pairwise anti-correlations.

As an aside, we apply the DEM formalism to a simple distribution, that is $\mathbf{x} \sim$ Categorical$([1 - p, p/3, p/3, p/3])$.
This should be recognizable as the probability distribution for a uniformly depolarizing single qubit error where we map $\{00, 10, 01, 11\} \to \{I, X, Z, Y\}$.
Proceeding through the mathematical steps outlined above yields
\begin{align*}
  \boldsymbol{\theta}
  &=
  \begin{pmatrix}
    \frac{1}{2} - \frac{1}{2\left(1 - 4p/3\right)^{3/2}} \\
    \frac{1}{2} - \frac{1}{2}\sqrt{1 - \frac{4}{3}p} \\
    \frac{1}{2} - \frac{1}{2}\sqrt{1 - \frac{4}{3}p} \\
    \frac{1}{2} - \frac{1}{2}\sqrt{1 - \frac{4}{3}p} \\
  \end{pmatrix}.
\end{align*}
The entries $\frac{1}{2} - \frac{1}{2}\sqrt{1 - \frac{4}{3}p}$ are identical to the value produced as a solution to \emph{independent} error processes representing uniform depolarizing single qubit errors in \cite{Gidney2020decorrelated-depolarization} and used widely in \texttt{stim}.

Consider next the probability distribution, $\mathbf{p} = [0.8, 0.1, 0.1, 0]^T$, which as input to \autoref{eq:excitation-rates-from-probability} yields excitation rate vector
\begin{align*}
  \boldsymbol{\theta} &\approx
  \begin{pmatrix}
    -0.308 \\
    0.113 \\
    0.113 \\
    -0.016
  \end{pmatrix}.
\end{align*}
Recalling our definitions of the elements of $\boldsymbol{\theta}$ (excluding the leading term, as discussed above) as Bernoulli parameters, we have an nonphysical parameter: $\theta_{11} < 0$.

%% Subsection: Further Reading
\subsection{Further Reading}
\label{subsec:further-reading}
The DEM parameterization of probability distributions on binary vectors is \emph{not} the only parameterization to interleave linear transforms with logarithms of (sometimes transformed) probabilities.
In this section we identify some similar, alternative methods in the literature.

The well studied class of \emph{log-linear} models for multivariate binary distributions is defined by \cite{wasserman2003statistics} to satisfy:
\begin{align}
  \ln p_X = \sum_{A \subseteq [n]} \lambda_A(X)
\end{align}
such that
\begin{enumerate}
\item $\lambda_\emptyset(X)$ is a constant.
\item $\lambda_A(X)$ is a function of \emph{only} the bits in $A$: $\lambda_A(X) = \lambda_A(A \cap X)$.
\item $\lambda_A(X)$ is zero when \emph{any} of these bits is zero: $\lambda_A(X) = \lambda_A(X)[X\cap A = A]$.
\end{enumerate}
Note that the final item implies
\begin{align}
  \ln p_X = \sum_{X \subseteq A \subseteq [n]} \lambda_A.
\end{align}
In matrix form, we have $\ln \mathbf{p} = \mathbf{Z}\boldsymbol{\lambda}$.
Note that $\mathbf{Z}$ here is the same as the one introduced in \ref{subsec:moments-and-more}.
This implies the relationship
\begin{align}
  \boldsymbol{\lambda} = \mathbf{Z}^{-1} \ln \mathbf{p}.
\end{align}
Note that in log-linear models $\lambda_A = 0$, if and only if there exists a partition of $A = \{A_1, A_2\}$ such that bits in $A_1$ are conditionally independent of bits in $A_2$ given the bits in $[n] - A$.
Additionally, for a subclass of such models---chordal, graphical models---there is an efficient algorithm for structure learning \cite{Petitjean2015-SDM}.

The class of \emph{log-mean linear} models introduced in \cite{2013_log-mean_linear_models_for_binary_data} is a parameterization of discrete binary variable distributions defined by the parameter vector
\begin{align}
  \boldsymbol{\gamma} = \mathbf{Z}^{-T} \ln \left( \mathbf{Zp} \right).
\end{align}
In that work, the authors claim given two disjoint sets $0 \subset A, B \subset [n]$ the bits in $A$ are \emph{independent} of the bits in $B$ if $\gamma_{A^{\prime}\cup B^{\prime}} = 0$ for all $0\subset A^{\prime}\subseteq A$ and $0\subset B^{\prime}\subseteq B$.
That work also notes other extant transforms of similar style.
Note that both the log-linear and log-mean linear models are \emph{complete} parameterizations of all probability distributions with strictly positive probabilities for each outcome.
Each model provides a smooth invertible transform from the bulk of the probability simplex.

Ignoring a constant factor, we have examined what might be called a \emph{log-polarization linear} model for probability distributions
\begin{align}
\boldsymbol{\psi} = \mathbf{H}^{-1} \ln \left(\mathbf{Hp}\right).
\end{align}
However, \cite{hendy1993} chose to call it the \emph{Hadamard conjugation} transform when using this to perform a similar estimation problem on phylogenetic trees based on observed gene frequencies.
Unlike the other two models above, we must account for the pole when $\mathbf{Hp}$ has a component equal to zero.
While the authors know of no general condition to avoid this, when $p_0 > \sum_{i=1}^{2^n - 1} p_i$ and all $p_i$ are non-zero there are no divergences.
Whether this presents a hurdle to (non-physical) log-polarization linear models providing a complete description of binary string distributions remains to be seen.
However, there is a very clear interpretation of the meaning of zeros in $\boldsymbol{\psi}$ (at least in the physical case); there is \emph{no} process which flips \emph{exactly} the bits indicated by that component.

%%%%%%%%%%%%%%
% ALGORITHMS %
%%%%%%%%%%%%%%
\section{DEM Estimation}
\label{sec:dem-estimation}
In this section, we discuss several algorithms for estimating detector error models from syndrome data.
We now assume a data-set, $\mathbf{X} \in \mathbb{F}_2^{n \times N}$, consisting of $N$ independent, identically distributed bit-vectors, $\mathbf{x}_i$, generated by a detector error model.
We will discuss two general classes of algorithms: rate estimation and structure learning.

%% Subsection: Moment-Based Algorithms
\subsection{Moment-Based Algorithms}
\label{subsec:moment-based-algorithm}
We start from \autoref{eq:neighborhood-moment} and note that we are attempting to estimate parameters which can be indexed by the columns of the incidence matrix $\mathbf{M}$.
One may simply compute the moments for \emph{exactly} the columns of the incidence matrix and then apply a numerical solver.
We do this with an additional approximation.

For a given $S \in \mathbf{M}$, we construct the primed incidence matrix and excitation consisting
only of $A \in \mathbf{M}$ with $\emptyset \ne A \cap S$.
We are interested in summing over only those terms which satisfy
\begin{align}
  \mathbf{M}^{\prime}\mathbf{e}^{\prime} = \mathbf{1}. \label{eq:primed-excitation-constraint}
\end{align}
We can form the augmented matrix and row reduce to find
\begin{align*}
  \left[ \mathbf{M}^{\prime} |  \mathbf{1} \right] \to [\mathbf{I}_{|S|} |\mathbf{F}| \mathbf{y}].
\end{align*}
Thus, for any vector $\mathbf{e}_f \in \mathbb{F}_2^{|\mathcal{N}(S)| - |S|}$, the vector
\begin{align*}
  \mathbf{e}^\prime
  =
  \begin{bmatrix}
    \mathbf{e}_d \\
    \hline
    \mathbf{e}_f
  \end{bmatrix}
  =
  \begin{bmatrix}
    \mathbf{Fe}_f + \mathbf{y} \\
    \hline
    \mathbf{e}_f
  \end{bmatrix}
\end{align*}
satisfies \autoref{eq:primed-excitation-constraint}.
We label the components of the primed excitation $\mathbf{e}_d$ for the dependent bits and $\mathbf{e}_f$ for the freely varying bits.
We then have a mildly more efficient, but exponential summation to calculate the moment predicted by a DEM:
\begin{align}
  \mu_S
  &= \sum_{\mathbf{e}_f \in \mathbb{F}_2^{|\mathcal{N}(S)| - |S|}}
  \Pr(\mathbf{e}_d = \mathbf{Fe}_f + \mathbf{y} | \boldsymbol{\theta})
  \Pr(\mathbf{e}_f | \boldsymbol{\theta}).
\end{align}
Finally, we recall the probability of an excitation scales as $\mathcal{O}(\langle \mathbf{\theta} \rangle^{|\mathbf{e}|})$ if all rates are approximately equal.
As we expect small rates, this vanishes quickly.
So we approximate with low-weight, free excitations
\begin{align}
  \label{eq:low-weight-approximate-moment}
  \tilde{\mu}_\mathbf{s}(\boldsymbol{\theta})
  &\approx \sum_{\mathbf{e}_f : |\mathbf{e}_f| \le w_{\max}}
  \Pr(\mathbf{e}_d = \mathbf{Fe}_f + \mathbf{y} | \boldsymbol{\theta})
  \Pr(\mathbf{e}_f | \boldsymbol{\theta}).
\end{align}
There is a researcher choice of which excitations will be free.
Choosing these to be the ones with the \emph{lowest} excitation rates was observed to reduce the variance of the estimates.
However, this cannot be done when the rates are not known.
For generality and predictability, we sort hyperedges by inclusion and use this (sub-optimal) ordering to partition excitations into dependent and free.
This detail is elided in the pseudo-code below.

Meanwhile, we can estimate the moment from data and do so using the posterior mean from a $\beta(1, 1)$ prior,
\begin{align}
  \label{eq:posterior-mean-moment-estimator}
  \widehat{\mu}_\mathbf{s} &= \frac{1}{2 + N}\left(1 + \sum_{\mathbf{x} \in \mathbf{X}} [\mathbf{x} \ge \mathbf{s}] \right).
\end{align}
We use this approach as it softens the numerical impact of moments with zero or one satisfying syndrome.
We mix statistical philosophies and construct the sample standard deviation,
\begin{align}
  \label{eq:moment-std-estimator}
  \sigma_\mathbf{s} &= \sqrt{\frac{\widehat{\mu}_\mathbf{s}(1 - \widehat{\mu}_\mathbf{s})}{N}}.
\end{align}
We complete our algorithm by setting the up the system of equations,
\begin{align}
  \label{eq:normalized-moment-residuals}
  r_{\mathbf{s}}(\mathbf{\theta}) = \frac{\tilde{\mu}_\mathbf{s}(\mathbf{\theta}) - \widehat{\mu}_\mathbf{s}}{\sigma_\mathbf{s}} = 0,
\end{align}
and solving using a root-finder, such as the options available in \texttt{scipy.optimize} \cite{2020SciPy-NMeth}.

\begin{algorithm}
  \caption{Estimate DEM parameters from moments.}
  \begin{algorithmic}
    \Function{EstimateFromMoments}{$\mathbf{X}, \mathbf{M}, w$}
    \State \textbf{Input:} $\mathbf{X}$ \Comment{Syndrome data-set.}
    \State \textbf{Input:} $\mathbf{M}$ \Comment{DEM incidence matrix.}
    \State \textbf{Input:} $w$ \Comment{Maximum weight of excitation to consider. Recommend $w \in \{2,3\}$.}
    \For{$\mathbf{s} \in \mathbf{M}$} \Comment{Pre-processing, can be parallel}
    \State Calculate $(\widehat{\mu}_\mathbf{s}, \sigma_\mathbf{s})$.
    \Comment{Eqs. \eqref{eq:posterior-mean-moment-estimator}, \eqref{eq:moment-std-estimator}}
    \State Compute reduced incidence matrix $\mathbf{M}^\prime$.
    \Comment{Eq. \eqref{eq:neighborhood-moment}}
    \State Row-reduce $[\mathbf{M}^\prime|\mathbf{1}]$ to find $(\mathbf{F}, \mathbf{y})$.
    \State Compute and store all valid $\mathbf{e}^{\prime} = [(\mathbf{Fe}_f)^{T} + \mathbf{y}^{T}|\mathbf{e}_f^{T}]^T$
    with $|\mathbf{e}_f| \le w$.
    \EndFor
    \State $\widehat{\boldsymbol{\theta}} \gets \widehat{\boldsymbol{\mu}}$
    \Comment{Initialize root-finding with linear guess.}
    \Repeat \Comment{Estimation via optimization}
    \State Compute $\widetilde{\boldsymbol{\mu}}(\widehat{\boldsymbol{\theta}})$.
    \Comment Eq. \eqref{eq:low-weight-approximate-moment}.
    \State Compute $\mathbf{r}(\widehat{\boldsymbol{\theta}})$.
    \Comment Eq. \eqref{eq:normalized-moment-residuals}.
    \State $\widehat{\boldsymbol{\theta}} \gets \widehat{\boldsymbol{\theta}}^{new}$ from root-finder.
    \Until $\widehat{\boldsymbol{\theta}}$ converges.
    \State \Return $\widehat{\boldsymbol{\theta}}$
    \EndFunction
  \end{algorithmic}
  \label{alg:estimate-parameters-moments}
\end{algorithm}

The main driver for the complexity of Algorithm \autoref{alg:estimate-parameters-moments} is the size of the primed excitation vectors computed in the for-loop and the repeat block to estimate $\widetilde{\boldsymbol{\mu}}(\widehat{\boldsymbol{\theta}})$.
In a sparse encoding, the size of these vectors goes as the number of non-zero entries, $|\mathbf e^\prime| = |\mathbf{e}_d| + |\mathbf{e}_f|$.
By construction, $|\mathbf{e}_f| \le w$, and $|\mathbf{e}_d| \le |S|$.
If $k = \max_{S\in\mathbf{M}} |S|$ is the maximum hyperedge cardinality, then $|\mathbf{e}^\prime| \le k + w$.

For hyperedge $S \in \mathbf{M}$, the number of $\mathbf{e}^\prime$ is the number of $\mathbf{e}_f$ for which $|\mathbf{e}_f| \le w$, which is $O(E_{S}^w)$, where $E_{S} = |\mathcal{N}(S)| - |S|$ is the dimension of $\mathbf{e}_f$.
The size of the neighborhood of $S$ can be approximated as $|\mathcal{N}(S)| \approx |S| E / n$, noticing that each detector in $S$ is included in an average of $E / n$ hyperedges.
Putting the pieces together, the total complexity of Algorithm \autoref{alg:estimate-parameters-moments} is
\begin{align}
  \mathcal{O}\left( E (k + w) \left(\frac{kE}{n}\right)^{w} \right)
  = \mathcal{O}\left( \frac{(kE)^{w + 1}}{n^{w}} \right).
\end{align}
The algorithm scales exponentially with $w$, but we present evidence in the Appendix \ref{sec:maximum-weight-in-moment-based} that $w \in \{ 2, 3 \}$ is sufficient.
At first glance, the complexity appears to decrease as the size of the syndrome increases.
However, for most realistic DEMs, $E$ is at least linear in $n$, so the net exponent on $n$ is positive.

We also may wish to learn the structure of $\mathbf{M}$.
We present Algorithm \autoref{alg:structure-learn-moments}, a moment-based algorithm for doing this in a limited case of low maximal hyperedge cardinality.
To begin, this algorithm finds statistically significant correlations from among all pairs of detectors using the analysis introduced in \cite{https://doi.org/10.1002/qute.201800012} and popularized in \cite{google2021exponentialsuppression}.
Here, we reproduce Equations 11 and 19 from the Supplementary Material of \cite{google2021exponentialsuppression} in our notation with minor algebraic changes:
\begin{align}
  \theta_{\{i,j\}}
  &= \frac{1}{2} - \frac{1}{2}\sqrt{
    \frac{\left(1 - 2\widehat{\mu}_{\{i\}}\right) \left(1 - 2\widehat{\mu}_{\{j\}}\right)}
         {1 - 2(\widehat{\mu}_{\{i\}} + \widehat{\mu}_{\{j\}} - 2\widehat{\mu}_{\{i, j\}})}
  }, \label{eq:theta_ij} \\
  \sigma_{\{i,j\}}
  &\approx \frac{1}{\sqrt{N}} \sqrt{
    \theta_{\{i,j\}}(1 - \theta_{\{i,j\}})
    + \frac{\widehat{\mu}_{\{i\}} \widehat{\mu}_{\{j\}} (1-\widehat{\mu}_{\{i\}}) (1-\widehat{\mu}_{\{j\}})}
    {(1-2\widehat{\mu}_{\{i\}})^2 (1-2\widehat{\mu}_{\{j\}})^2}
  }. \label{eq:sigma_ij}
\end{align}

For sufficiently large samples, $\theta_{\{i, j\}}$ is approximately normally distributed around the true value with standard deviation $\sigma_{\{i, j\}}$.
In the absence of correlations, the $z$-score $\theta_{\{i, j\}} / \sigma_{\{i, j\}}$ is approximately standard normal, and the most extreme value is expected to be $\Phi^{-1}(1 - \binom{n}{2}^{-1})$, where $\Phi^{-1}$ is the standard normal quantile function.
We therefore define a significant pairwise correlation as any $(i,j)$ whose $z$-score exceeds this threshold.

The \emph{correlation graph}, $\mathcal{G}$, whose edges comprise all statistically significant pairwise correlations, is a useful starting point.
Any excitation will induce correlations between all pairs of detectors in the corresponding hyperedge.
Therefore, the search space for DEM hyperedges is restricted to cliques of $\mathcal{G}$.

The algorithm proceeds through this search space by iteratively growing a frontier of order-$k$ hyperedges, augmenting each hyperedge with a single additional detector and testing whether the resulting hyperedge of order $k+1$ is a clique in $\mathcal{G}$ and has a significant residual moment.
Termination occurs either at a specified $k_{\max}$ or when no more significant cliques are found.
The algorithm has two varieties, depending on how it is initialized.
By default, the initial frontier comprises all singleton detectors, and the search can in principle find any significant hyperedge up to the maximum $k$.
This form is general but costly and produces large DEMs which can be difficult to interpret.
If the user provides a seed set of hyperedges, then the algorithm will return a restricted DEM, in which all hyperedges are supersets of one or more seeds.
This targeted variant is useful in a two-stage exploratory workflow: first, $\theta_{\{i,j\}}$ analysis identifies significant and unexpected correlations which warrant explanation; these correlations are then used individually as seeds to grow a restricted DEM that identifies any higher-order hyperedges that contribute to the pairwise correlation.
Section \ref{subsubsec:correlated-measurement-error} presents an example of this workflow applied to Google's QEC data.

\begin{algorithm}
  \caption{Strucuture learning from moments.}
  \label{alg:structure-learn-moments}
  \begin{algorithmic}
    \Function{LearnFromMoments}{$\mathbf{X}, w_{search}, w_{fit}, k_{\max}, \mathcal{F}=\{\{i\} \text{ for } i=1,\cdots,n\}$}
    \State \textbf{Input:} $\mathbf{X}$
    \Comment{Syndrome data-set.}
    \State \textbf{Input:} $w_{search}$
    \Comment{Controls fidelity of moment approximation during structure discovery. Recommend $w_{search}=2$}
    \State \textbf{Input:} $w_{fit}$
    \Comment{Controls fidelity of moment approximation for final rate estimation. Recommend $w_{fit}=3$}
    \State \textbf{Input:} $k_{\max}$
    \Comment{Maximum hyperedge cardinality to consider.}
    \State \textbf{Input (optional):} $\mathcal{F}$
    \Comment{Seed hyperedges. Must have equal cardinality. Default: all singletons.}
    \State $\mathcal{G} \gets \lbrace (i, j) \colon \theta_{\lbrace i, j \rbrace}/\sigma_{\lbrace i, j \rbrace} > \Phi^{-1}(1 - \binom{n}{2}^{-1}) \rbrace$
    \Comment{Eqs. \eqref{eq:theta_ij}, \eqref{eq:sigma_ij}}
    \State $\mathcal{D} \gets \mathcal{F}$
    \Comment{DEM hyperedges}
    \State $k \gets$ cardinality of seeds
    \While{$k < k_{\max}$}
    \State $\mathcal{F}^\prime \gets \emptyset$
    \Comment{Frontier hyperedges of cardinality $k+1$}
    \For{$S \in \mathcal{F}$}
    \For{$i \in [n] \setminus S$}
    \If{$(i,j) \in \mathcal{G}$ for all $j \in S$}
    \State $\mathcal{F}^\prime \gets \mathcal{F}^\prime \cup (\{i\} \cup S)$
    \Comment{$(\{i\} \cup S)$ is a $(k+1)$-clique in $\mathcal G$}
    \EndIf
    \EndFor
    \EndFor
    \State $\boldsymbol{\widehat{\theta}} \gets \Call{EstimateFromMoments}{\mathbf{X}, \mathcal{D}, w_{search}}$
    \Comment Alg. \ref{alg:estimate-parameters-moments}
    \State Approximate $\widetilde{\mu}_S(\boldsymbol{\widehat{\theta}})$ for all $S \in \mathcal{F}^\prime$
    \Comment{Eq. \eqref{eq:low-weight-approximate-moment}}
    \State Estimate $(\widehat{\mu}_S, \sigma_S)$ for all  $S \in \mathcal{F}^\prime$
    \Comment{Eqs. \eqref{eq:posterior-mean-moment-estimator}, \eqref{eq:moment-std-estimator}}
    \State Compute $\mathbf{r} = (\boldsymbol{\widehat{\mu} -\widetilde{\mu}}) / \boldsymbol{\sigma}$.
    \Comment{Eq. \eqref{eq:normalized-moment-residuals}}
    \State $
    \mathcal{F}^\prime \gets \left\{ S \in \mathcal{F}^\prime \colon r_S > \Phi^{-1}\left(1 - \binom{n}{k+1}^{-1} \right) \right\}
    $
    \Comment{New hyperedges are significantly non-normal in residual space.}
    \If{$\mathcal{F}^\prime = \emptyset$}
    \State \textbf{break}
    \EndIf
    \State $\mathcal{D} \gets \mathcal{F}^\prime \cap \mathcal{D}$
    \State $\mathcal{F} \gets \mathcal{F}^\prime$
    \State $k \gets k+1$
    \EndWhile
    \State $\boldsymbol{\widehat{\theta}} \gets \Call{EstimateFromMoments}{\mathbf{X},\mathcal{D},w_{fit}}$
    \For{$S \in \mathcal D$}
    \If{$\widehat{\theta}_S / \sigma_S < \Phi^{-1}(1 - 1/|\mathcal D|)$}
    \State Delete $S$ from $\mathcal D$ and $\widehat{\theta}_S$ from $\boldsymbol{\widehat{\theta}}$
    \Comment{Prune insignificant hyperedges and corresponding rates}
    \EndIf
    \EndFor
    \State \Return $\mathcal D, \boldsymbol{\widehat{\theta}}$
    \EndFunction
  \end{algorithmic}
\end{algorithm}

%% Subsection: Parity Based Algorithms
\subsection{Parity Based Algorithms}
\label{subsec:parity-based-algorithms}
Two works \cite{remm2025, blumekohout2025estimatingdetectorerrormodels} have recently suggested algorithms for estimating DEMs from syndromes via parities.
While \cite{remm2025} does not make their estimation algorithm explicit, we believe it is equivalent to that presented in Section 4.1 of \cite{blumekohout2025estimatingdetectorerrormodels}.
We demonstrate our reasoning below before providing pseudo-code.
Return to \autoref{eq:GZpsi-is-Lomega_sums}, but now assume that $S$ has the highest cardinality of any hyperedge in the DEM, i.e. $S \subset A \implies \psi_A = 0$, resulting in
\begin{align}
  \psi_{S^+} \equiv \sum_{S \subseteq A \subseteq [n]} \psi_A
  &= -\frac{2}{2^{|S|}} \sum_{B \subseteq S} (-1)^{|B|} \omega_B, \label{eq:aggregated-attenuation} \\
  \psi_S
  &= -\frac{2}{2^{|S|}} \sum_{B \subseteq S} (-1)^{|B|} \omega_B.
\end{align}
We note that $\psi_{S^+}$ is a special case of \emph{aggregated attenuation}, where the sum ranges over the DEM hyperedges flipping \emph{all} detectors in $S$.

To compute the attenuation of all hyperedges in a DEM, one begins with \autoref{eq:aggregated-attenuation}  for all hyperedges of maximum cardinality. Armed with $\psi_S$ for maximal hyperedges, one iterates in order of decreasing $|S|$ to calculate
\begin{align}
  \psi_{S} = -\frac{2}{2^{|S|}} \sum_{B \subseteq S} (-1)^{|B|} \omega_B - \sum_{S \subset A} \psi_A.
\end{align}
using attenuations from previous iterations to compute the second summation in the RHS.
A logarithm and some manipulation changes this to exactly Equation 10 from \cite{remm2025} as shown in \eqref{eq:remm-equation-10}.

If $k = \max_{S\in\mathbf{M}} |S|$, then in the worst case, this algorithm requires the calculation of up to $\sum_{j=1}^k \binom{n}{j}$ depolarizations.
Similar to the algorithms in Section \ref{subsec:moment-based-algorithm}, this algorithm is prohibitively expensive unless the DEM has a relatively low $k$ and such hyperedges appear relatively rarely.
Happily, DEMs for surface codes appear to satisfy these conditions.
In such cases, we recommend the use of Algorithm \ref{alg:estimate-parameters-parities}.

\begin{algorithm}
  \caption{Estimate DEM parameters from parities.}
  \label{alg:estimate-parameters-parities}
  \begin{algorithmic}
    \Function{EstimateFromParities}{$\mathbf{X}, \mathbf{M}$}
    \State \textbf{Input:} $\mathbf{X}$
    \Comment{Syndrome data-set.}
    \State \textbf{Input:} $\mathbf{M}$
    \Comment{DEM incidence matrix.}
    \State Initialize empty map \texttt{supersets\_of}
    \Comment{$\mathcal{P}[n] \to$ subset of $\mathcal{P}[n]$}
    \State Initialize empty map $\boldsymbol{\widehat{\omega}}$
    \Comment{Estimated Depolarizations. $\mathcal{P}[n]\to\mathbb{R}$}
    \State Initialize empty map $\boldsymbol{\widehat{\psi}}$
    \Comment{Estimated Attenuations. $\mathcal{P}[n] \to \mathbb{R}$}
    \State Sort columns of $\mathbf{M}$ by decreasing hyperedge cardinality.
    \Comment{Larger sets now come first.}
    \For{$S \in \mathbf{M}$}
    \State $\boldsymbol{\widehat{\psi}}[S] \gets 0$
    \For{$B \subseteq S$}
    \If{$B \neq S$}
    \State \texttt{supersets\_of}$[B] \gets \{S\} \; \cap$ \texttt{supersets\_of}[$B$]
    \EndIf
    \If{$B \notin \boldsymbol{\widehat{\omega}}$}
    \State $\boldsymbol{\widehat{\omega}}[B] \gets -\ln \left[1 - 2 \frac{1}{N + 2} \left(1 + \sum_{\mathbf{x}\in \mathbf{X}} \mathbf{x}\cdot\mathbf{b} \right)\right]$
    \Comment{Estimated depolarization from syndromes with a $\beta(1,1)$ prior.}
    \EndIf
    \State $\boldsymbol{\widehat{\psi}}[S] \gets \boldsymbol{\widehat{\psi}}[S] + (-1)^{|B|}\boldsymbol{\widehat{\omega}}[B]$
    \EndFor
    \State $\boldsymbol{\widehat{\psi}}[S] \gets -\frac{2 \boldsymbol{\widehat{\psi}}[S]}{2^{|S|}}$
    \For{$A \in$ \texttt{supersets\_of}[$S$]}
    \State $\boldsymbol{\widehat{\psi}}[S] \gets \boldsymbol{\widehat{\psi}}[S] - \boldsymbol{\widehat{\psi}}[A]$
    \EndFor
    \EndFor
    \State \Return $\widehat{\mathbf{\psi}}$
    \EndFunction
  \end{algorithmic}
\end{algorithm}

To perform structure learning, we present Algorithm \ref{alg:structure-learn-parities}, noting the same requirement for maximal hyperedge cardinality to be relatively low.
This algorithm is structurally similar to Algorithm \ref{alg:structure-learn-moments}.
Both structure learning algorithms are likely to benefit from better hypothesis testing to prune hyperedges as in \cite{10.1145/2939672.2932775}.

One difference between the two algorithms is that Algorithm \ref{alg:structure-learn-parities} adjudicates frontier hyperedges on the basis of aggregated attenuation, rather than estimated rate.
To evaluate statistical significance, it must transform the estimated standard deviation of the rate, \autoref{eq:moment-std-estimator}, into an estimated standard deviation for the aggregated attenuation.
Recalling the relationship between attenuation and rate in \autoref{eq:rate-to-attenuation} and taking the derivative yields $d\psi/d\theta = 2/(1 - 2\theta) \approx 2$ for small $\theta$.
Propagation of uncertainty through this derivative leads to the $2\sigma_S$ in the denominator of the expression used to evaluate statistical significance of frontier hyperedges in Algorithm \ref{alg:structure-learn-parities}.

\begin{algorithm}
  \caption{Structure learning from parities.}
  \label{alg:structure-learn-parities}
  \begin{algorithmic}
    \Function{LearnFromParities}{$\mathbf{X}, k_{\max}, \mathcal{F}$}
    \State \textbf{Input:} $\mathbf{X}$
    \Comment{Syndrome data-set.}
    \State \textbf{Input:} $k_{\max}$
    \Comment{Maximum hyperedge cardinality to consider.}
    \State \textbf{Input (optional):} $\mathcal{F}$
    \Comment{Seed hyperedges from which to grow. Must have equal cardinality. Default: Singletons.}
    \State $\mathcal{G} \gets \{ (i, j) : \theta_{\{i,j\}}/\sigma_{\{i,j\}} > \Phi^{-1}(1 - \binom{n}{2}^{-1}) \}$
    \Comment{Eqs. \eqref{eq:theta_ij}, \eqref{eq:sigma_ij}}
    \State $\mathcal{D} \gets \mathcal{F}$
    \Comment{DEM hyperedges}
    \State $k \gets$ cardinality of seeds
    \While{$k < k_{\max}$}
    \State $\mathcal{F}^\prime \gets \emptyset$
    \Comment{Frontier hyperedges of cardinality $k+1$}
    \For{$S \in \mathcal{F}$}
    \For{$i \in [n] \setminus S$}
    \If{$(i,j) \in \mathcal{G}$ for all $j \in S$}
    \State $\mathcal{F}^\prime \gets \mathcal{F}^\prime \cup (\{i\} \cup S)$
    \Comment{$(\{i\} \cup H)$ is a $(k+1)$-clique in $\mathcal G$}
    \EndIf
    \EndFor
    \EndFor
    \State Compute $\psi_{S^+} = -\frac{2}{2^{|S|}} \sum_{B \subseteq S} (-1)^{|B|} \omega_B$
    \Comment{Eq. \eqref{eq:aggregated-attenuation}}
    \State Compute $\sigma_S$ using moment
    \Comment{Eq. \eqref{eq:moment-std-estimator}}
    \State $\mathcal{F}^\prime \gets \left\{ S \in \mathcal{F}^\prime \colon \psi_{S^+}/(2\sigma_S) > \Phi^{-1}\left(1 - \binom{n}{k+1}^{-1} \right) \right\}$
    \Comment{New hyperedges have significant aggregated attenuation.}
    \If{$\mathcal{F}^\prime = \emptyset$}
    \State \textbf{break}
    \EndIf
    \State $\mathcal{D} \gets \mathcal{F}^\prime \cap \mathcal{D}$
    \State $\mathcal{F} \gets \mathcal{F}^\prime$
    \State $k \gets k+1$
    \EndWhile
    \State $\boldsymbol{\widehat{\psi}} \gets \Call{EstimateFromParities}{\mathbf{X},\mathcal{D}}$
    \State $\boldsymbol{\widehat{\theta}} \gets (1 - e^{-\boldsymbol{\widehat{\psi}}})/2$
    \For{$S \in \mathcal D$}
    \If{$\widehat{\theta}_S / \sigma_S < \Phi^{-1}(1 - 1/|\mathcal D|)$}
    \State Delete $S$ from $\mathcal D$ and $\widehat{\theta}_S$ from $\boldsymbol{\widehat{\theta}}$
    \Comment{Prune insignificant hyperedges and corresponding rates.}
    \EndIf
    \EndFor
    \State \Return $\mathcal D, \boldsymbol{\widehat{\theta}}$
    \EndFunction
  \end{algorithmic}
\end{algorithm}

We also examine the possibility of rate estimation from parities from pseudo-inversion of a tall matrix a.k.a least-squares regression as suggested in \cite{blumekohout2025estimatingdetectorerrormodels}.
The benefit of this approach is an algorithm which is no longer exponential in the weight of hyperedges in the DEM.
In particular we must take at least $E$ distinct samples to reach a full-rank design matrix, after which the computational complexity is determined by the cost of least-squares estimation or pseudo-inverse calculation.
In some cases, this can be \emph{significantly} faster than Algorithm \ref{alg:estimate-parameters-parities}.
We note in the next section some subtleties that arise due to the noise experienced by depolarizations, which prevent these algorithms from working for all DEMs and place non-obvious constraints on the set of depolarizations one may wish to calculate.

\begin{algorithm}
  \caption{Estimate excitation rates from parities via least-squares.}
  \label{alg:estimate-from-parities-lsqr}
  \begin{algorithmic}
    \Function{EstimateFromParitiesLSQR}{$\mathbf{X}, \mathbf{M}, \mathbf{Y}$}
    \State \textbf{Input:} $\mathbf{X}$
    \Comment{Syndrome data-set.}
    \State \textbf{Input:} $\mathbf{M}$
    \Comment{DEM incidence matrix.}
    \State \textbf{Input:} $\mathbf{Y}$
    \Comment{Columns indicate detector subsets to query.}
    \State $\widehat{\boldsymbol{\omega}} \gets \mathbf{0}^{n_r}$
    \For{$i \in [n_r]$}
    \State $\widehat{\boldsymbol{\omega}}[i] \gets -\ln \left[1 - 2 \frac{1}{N + 2} \left(1 + \sum_{\mathbf{x}\in \mathbf{X}} \mathbf{y}_i\cdot\mathbf{x} \right)\right]$
    \Comment{$\mathbf{y}_i$ is $i$-th column of $\mathbf{Y}$}
    \EndFor
    \State $\mathbf{A} \gets \mathbf{Y}^T\mathbf{M}$
    \Comment{Math in $\mathbb{F}_2$.}
    \State $\widehat{\boldsymbol{\psi}} \gets \mathbf{A}^{+} \widehat{\boldsymbol{\omega}}$
    \Comment{Implicit from least-squares algorithm.}
    \State \Return $\widehat{\boldsymbol{\psi}}$
    \EndFunction
  \end{algorithmic}
\end{algorithm}

%% Subsection: Accuracy
\subsection{Accuracy}
\label{subsec:accuracy}
To evaluate the accuracy of these estimation algorithms, we begin with syndromes generated by known DEMs and compare the withheld true rates (and structure, as appropriate) against those estimated by the algorithms.
For the generating DEM, we use the SI1000-decorated \texttt{stim} circuits published in the Google data-set alongside the syndromes collected from hardware.
We first discuss the performance of structure learning algorithms followed by the convergence behavior of parameter estimation on both the SI1000 circuits and more general cases.

\begin{table}
  \centering
  \begin{tabular}{ccccccccc}
    \hline
     & & & \multicolumn{2}{c}{False Positives} & \multicolumn{2}{c}{False Negatives} & \multicolumn{2}{c}{Time (s)} \\
    Code & Dist. & Shots & Alg.~\ref{alg:structure-learn-moments} & Alg.~\ref{alg:structure-learn-parities} & Alg.~\ref{alg:structure-learn-moments} & Alg.~\ref{alg:structure-learn-parities} & Alg.~\ref{alg:structure-learn-moments} & Alg.~\ref{alg:structure-learn-parities} \\
    \hline
    Rep. &  9 & $10^6$ & 1 (0.5\%) & 0 (0.0\%) & 0 (0.0\%) & 0 (0.0\%) & 4.8 & 1.9 \\
    Rep. & 19 & $10^6$ & 0 (0.0\%) & 0 (0.0\%) & 0 (0.0\%) & 0 (0.0\%) & 18.6 & 3.2 \\
    Rep. & 29 & $10^6$ & 0 (0.0\%) & 0 (0.0\%) & 0 (0.0\%) & 0 (0.0\%) & 199.3 & 7.2 \\
    \hline
    Surf. & 3 & $10^6$ & 0 (0.0\%) &  0 (0.0\%) &  16 (10.3\%) &  29 (18.7\%) &   9.6 &  3.2 \\
    Surf. & 5 & $10^6$ & 0 (0.0\%) &  0 (0.0\%) & 102 (15.3\%) & 157 (23.5\%) &  42.3 &  3.7 \\
    Surf. & 7 & $10^6$ & 0 (0.0\%) & 18 (1.6\%) & 403 (26.3\%) & 456 (29.8\%) & 522.5 & 13.7 \\
    \hline
    Surf. & 3 & $10^7$ & 0 (0.0\%) & 0 (0.0\%) &  1 (0.6\%) &  0 (0.0\%) &  80.6 & 5.1 \\
    Surf. & 5 & $10^7$ & 0 (0.0\%) & 2 (0.3\%) & 18 (2.7\%) &  9 (1.3\%) &  55.5 & 9.9 \\
    Surf. & 7 & $10^7$ & 6 (0.4\%) & 0 (0.0\%) & 89 (5.8\%) & 50 (3.3\%) & 480.9 & 15.2 \\
    \hline
  \end{tabular}
  \caption{Performance of structure-learning algorithms on simulated data.
    For each given code and distance, a \texttt{stim} circuit was created, decorated with the SI1000 noise model and used to sample the indicated number of shots.
    Then, Algorithms \autoref{alg:structure-learn-moments} and \autoref{alg:structure-learn-parities} were used to generate candidate DEM excitations for comparison with the DEM excitations constructed along-side the \texttt{stim} circuit.
    Each row represents a single experiment.
    A false positive indicates that the learning algorithm identifies an excitation not present in \texttt{stim} and the parenthesized percentage indicates the fraction of learned excitations not in the DEM.
    A false negative means that the learning algorithm excludes an excitation present in \texttt{stim}'s DEM and the parenthesized percentage indicates the fraction of true DEM excitations not discovered.
    Note that timings combine calculation of statistics and fitting.}
  \label{tab:structure-learning-performance}
\end{table}

\autoref{tab:structure-learning-performance} compares Algorithms \ref{alg:structure-learn-moments} and \ref{alg:structure-learn-parities} on the task of structure learning from syndromes generated for surface and repetition codes of various distances.
In structure learning, a \emph{false positive} is a hyperedge returned by the algorithm that is not in the true DEM, whereas a \emph{false negative} refers to a hyperedge in the true DEM that was not found by the algorithm.
The two algorithms have qualitatively similar accuracy (run-time is discussed in the next section): both algorithms tend to minimize false positives at the expense of significant false negatives for higher-distance surface codes.
This tendency is good for exploratory and diagnostic scenarios, in which the experimenter prizes confidence in the meaningfulness of discovered hyperedges over complete coverage of the true DEM.
However, in situations where a different balance is desired, an additional input parameter could be used to tune the permissiveness of the filter for frontier hyperedges.
For both algorithms, increasing the number of shots from $10^6$ to $10^7$ for the surface codes reduced the false negative rates without significantly affecting the false positive rates.
As in most learning applications, more data is the surest path to better answers.

\begin{figure}
  \centering
  \includegraphics{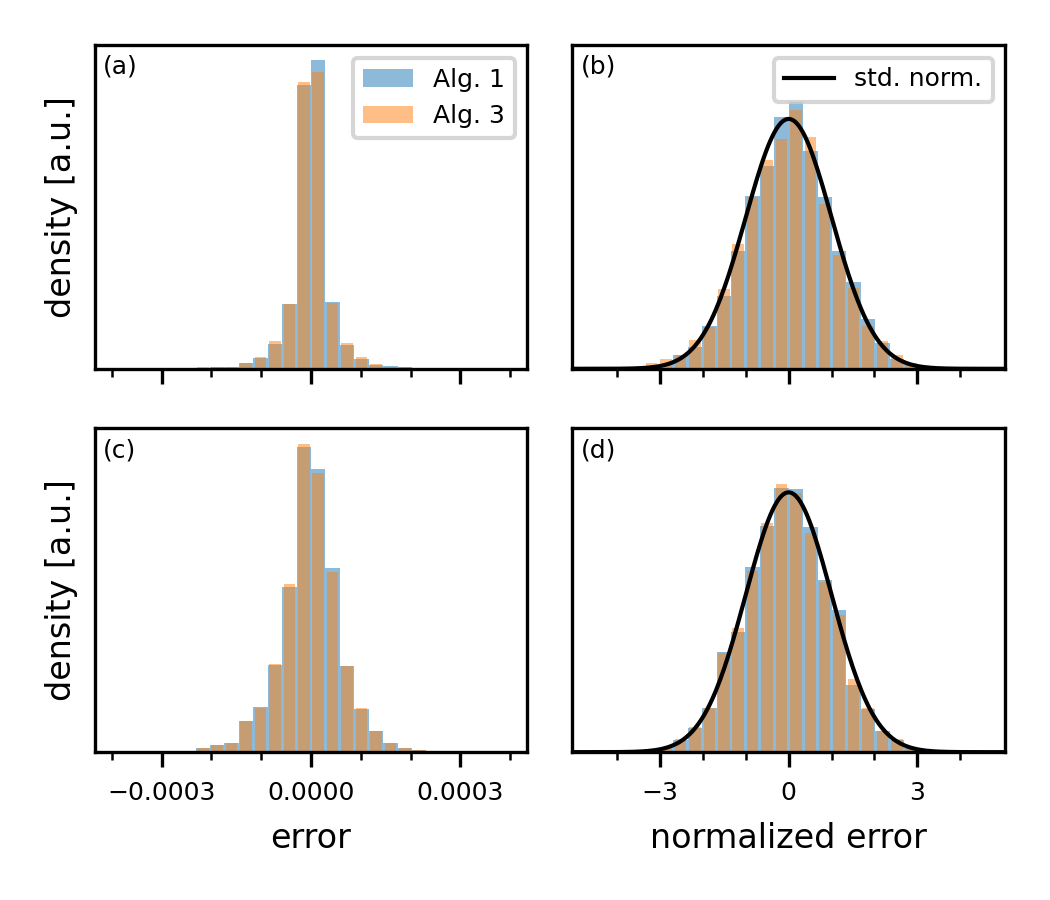}
  \caption{Residual errors from fitted and true values.
    The true parameters correspond to the SI1000 DEM for $d$ rounds from the temporal bulk of a $d=7$ surface code (top) or a $d=29$ repetition code (bottom).
    The SI1000 noise model was used in \texttt{stim} to produce $10^6$ syndromes.
    Then the parameters were estimated with the algorithms in the previous section.
    In all plots, blue corresponds to the moment-based Algorithm \ref{alg:estimate-parameters-moments} and orange represents the parameters estimated using the parity-based Algorithm \ref{alg:estimate-parameters-parities}.
    Histograms on the left show raw differences between estimated and true DEM parameters.
    Histograms on the right show normalized differences, wherein each error term has been divided by the approximation of standard error in \autoref{eq:moment-std-estimator}.
    The standard normal density function (black line) is superimposed on the normalized histograms and shows qualitative agreement.}
    \label{fig:fitted-residuals}
\end{figure}

On the task of rate estimation, both approaches perform similarly well.
\autoref{fig:fitted-residuals} shows raw and normalized (see \autoref{eq:moment-std-estimator} and \autoref{subsec:dem-estimate-variances}) errors between estimated and true rates for Algorithms \ref{alg:estimate-parameters-moments} and \ref{alg:estimate-parameters-parities}.
Rates were estimated from $10^6$ shots of syndromes generated by the SI1000 DEM for either a $d = 7$ surface code or a $d = 29$ repetition code.
On both codes, the two algorithms have indistinguishable performance.
Additionally, the normalized errors for both algorithms agree qualitatively with the standard normal distribution (black line), implying \autoref{eq:moment-std-estimator} is a good approximation for the standard error of the estimated rates.

\begin{figure}
  \centering
  \includegraphics{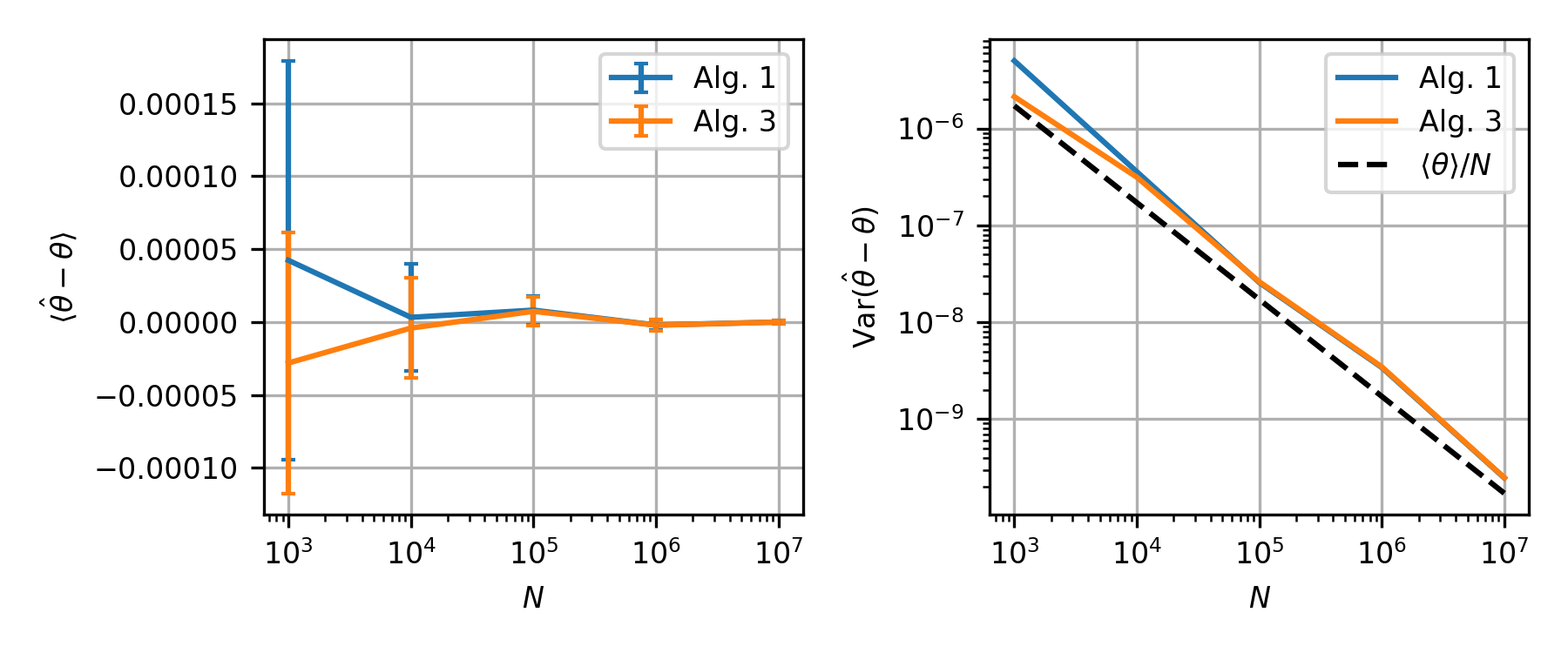}
  \caption{Bias (left) and variance (right) of estimated rates vs. number of samples (shots) from the SI1000 DEM for 3 rounds of a $d=3$ surface code.
    Rates were estimated from sampled syndromes via the moment-based Algorithm \ref{alg:estimate-parameters-moments} (blue) or the parity-based Algorithm \ref{alg:estimate-parameters-parities} (orange).
    In the left-hand plot, error bars denote the standard error of the mean, $\sqrt{\mathsf{Var}(\hat{\theta} - \theta)/E}$.
    In the right-hand plot, the function $\langle \theta \rangle / N$ is shown (black, dashed line) as a guide to the eye, where $\langle \theta \rangle$ is the mean of the true DEM rates.}
  \label{fig:variance-of-error}
\end{figure}

Estimation error is presumably a combination of systematic error---bias in the algorithm---and random error from finite numbers of shots.
We investigated the sample mean and variance of raw (un-normalized) estimation error as a function of the number of shots, and the results are shown in \autoref{fig:variance-of-error}.
The left-hand plot shows that estimates are unbiased, in that the mean difference between estimated and true rate is always within standard error of zero.
In the right-hand plot, the variance of the estimation error is proportional to $\theta/\sqrt{N}$, which is the expected scaling for random error, implying that estimation error is dominated by shot noise.

\begin{figure}
  \centering
  \includegraphics[width=0.5\textwidth]{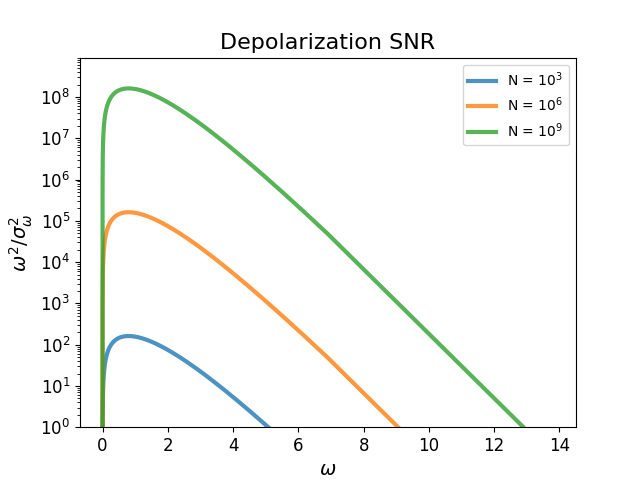}
  \caption{The SNR (\autoref{eq:depolarization-snr}) as a function of depolarization for different shot-counts.}
  \label{fig:snr-curves}
\end{figure}

In the context of parameter estimation, the behavior observed above may not apply to Algorithms \ref{alg:estimate-parameters-parities} and \ref{alg:estimate-from-parities-lsqr} in general.
To understand this potential systematic limit, we examine the signal-to-noise ratio (SNR) of depolarizations.
We express the predicted variance of a depolarization $\omega_S$ first as a function of the corresponding polarization and then the depolarization,
\begin{align}
  \sigma^2_{\omega_S}
  = \left(\frac{\partial \omega_S}{\partial \pi_S}\right)^2 \sigma^2_{\pi_S}
  = \frac{1 - \pi_S^2}{\pi_S^2}
  = e^{2\omega_S} - 1.
\end{align}
When estimated with a finite data-set of size $N$, we construct the SNR for a specific depolarization
\begin{align}
  \mathrm{SNR} = \frac{\widehat{\omega}_S^2}{\widehat{\sigma}^2_{\omega_S}} = \frac{N\widehat{\omega}_S^2}{e^{2\widehat{\omega}_S} - 1}. \label{eq:depolarization-snr}
\end{align}
Figure \ref{fig:snr-curves} shows this value for some values of $N$.
The key features to observe are that the SNR is maximized at $\omega_S \approx 0.797$ and that, for a reasonable number of shots, the SNR is only appreciably higher than one for a relatively small range of values.
Specifically, if one is only able to query depolarizations with a specific range of a central value $\overline{\omega}$, we require enough data to make the SNR well above one:
\begin{align}
  1 \ll \frac{N\overline{\omega}^2}{e^{2\overline{\omega}} - 1} \implies e^{2\overline{\omega}} \ll N\overline{\omega}^2 + 1
\end{align}
Thus our shot-count requirements are exponential in $\overline{\omega}$.

We further examine the requirements for the depolarization queries - denoted as $\mathbf{Y}$ in Algorithm \autoref{alg:estimate-from-parities-lsqr}.
Recall, for a given DEM,
\begin{align}
  \omega_S = \sum_{A \in \mathbf{M}} (|A\cap S| \bmod 2) \psi_A.
\end{align}
If we have $\psi_A \approx \overline{\psi}$ for all $A$ and the variance is not too great (i.e. everything is the same order of magnitude), then
\begin{align}
  \omega_S \approx \overline{\psi} \sum_{A \in \mathbf{M}} (|A\cap S| \bmod 2).
\end{align}
Which is to say $\omega_S \approx \overline{\psi} \times$ (number of odd-overlap hyperedges).
The problem of choosing depolarizations with which to estimate excitation rates via least-squares reduces to finding suitable $\mathbf{Y}$ such that the number of odd-overlap hyperedges lands you near the SNR peak and that the resulting system of equations is of full rank.
Mathematically, we must choose $\mathbf{Y}$ such that
\begin{align}
  \mathrm{rank} (\mathbf{Y}^T\mathbf{M}) = E \;\;\;\text{and}\;\;\; \frac{0.797}{\overline{\psi}} \approx \sum_{\mathbf{s \in M}} \mathbf{y \cdot s} \;\;\;\forall\;\mathbf{y \in Y}.
\end{align}
Note that with infinite shots, these algorithms do yield numerical precision.

There is at least one class of DEMs for which least-squares (or likely other methods of parity based estimation) will fail, uniformly-random DEMs.
A uniformly-random DEM over $n$-bits has hyperedges which are uniformly-randomly sampled from $\mathbb{F}_2^n$.
Let $u_\mathbf{y}$ be the number of odd-parity overlaps with bit-vector $\mathbf{y}$.
In a uniformly-random DEM, one can show that $u_\mathbf{y} \sim$ Binomial($E, 1/2$).
On average - every query touches half of the hyperedges.
Thus, if $E \gg 2 / \overline{\psi}$, we require exponentially many shots to form an estimate.

%% Subsection: DEM Estimation Timings
\subsection{DEM Estimation Timings}
\label{subsec:dem-estimation-timings}
\begin{figure}
  \centering
  \includegraphics{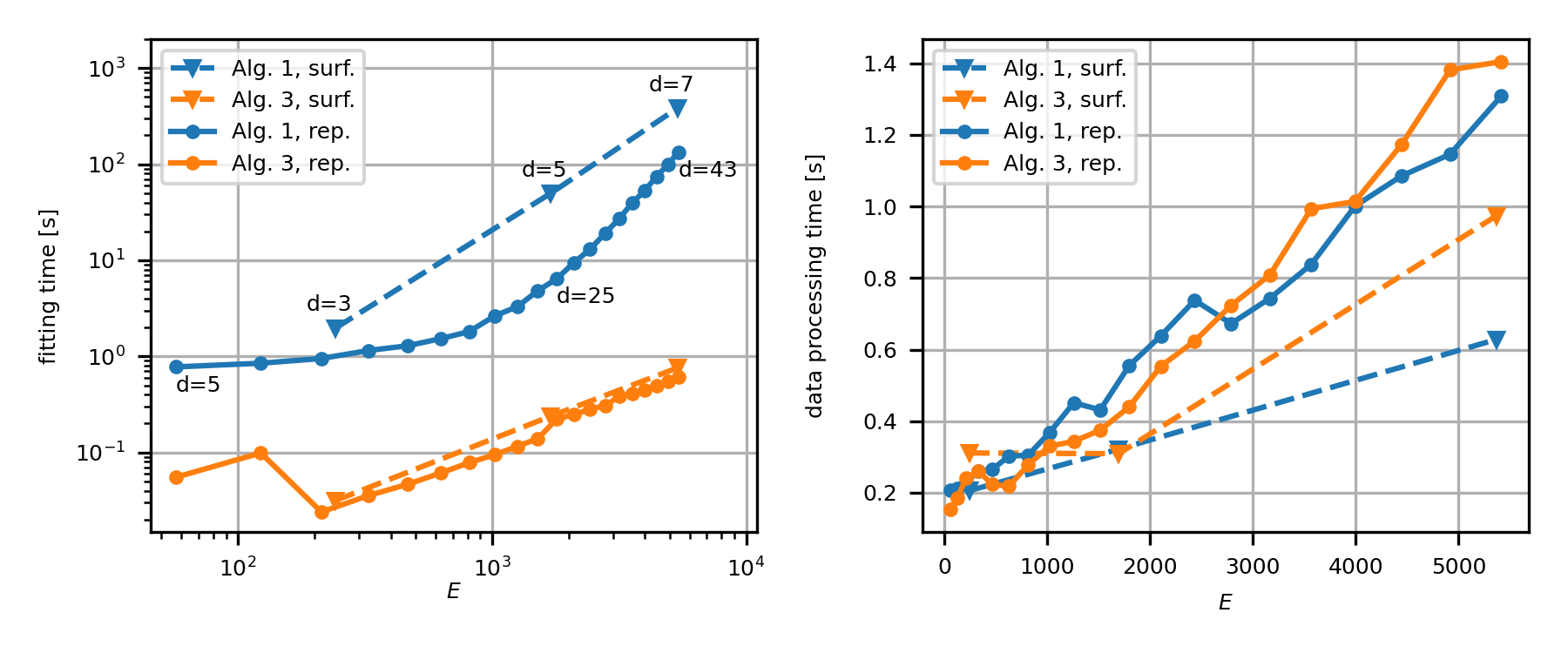}
  \caption{Scaling of rate estimation algorithms with DEM size. Solid lines and circles: repetition codes with $d \in \{5, 6, ..., 42, 43\}$.
    Dashed lines and triangles are $XZZX$-surface codes with $d \in \{3, 5, 7\}$.
    All syndrome data-sets had $10^6$ shots.
    Blue represents moment-based Algorithm \ref{alg:estimate-parameters-moments}, whereas orange denotes parity-based Algorithm \ref{alg:estimate-parameters-parities}.
    Left: Run-time of rate-estimation algorithms as a function of $E$, the number of hyperedges in the DEM.
    Right: Time required to calculate necessary statistics (moments or depolarizations) from syndromes vs $E$.}
  \label{fig:estimation-performance}
\end{figure}

We study the wall-time to execute the algorithms described in Sections \ref{subsec:parity-based-algorithms} and \ref{subsec:moment-based-algorithm}.
We do this for simulated data generated by \texttt{stim} using the SI1000 decorated versions of the $XZZX$-surface code and the repetition code.
In the run-times shown in \autoref{fig:estimation-performance}, we take the hyperedges of the SI1000 DEM and estimate parameters with a single sample of $10^6$ shots for each data point using Algorithms \ref{alg:estimate-parameters-moments} and \ref{alg:estimate-parameters-parities}.
In \autoref{tab:structure-learning-performance}, we demonstrate the performance of the Algorithms \ref{alg:structure-learn-moments} and \ref{alg:structure-learn-parities}.
All algorithms exploit parallelism, especially when computing statistics (moments or depolarizations) from syndromes, and all timings were measured on a server with 112 CPUs and 1 TB memory.

For both codes, the maximum cardinality of hyperedges in SI1000 is small - $k=2$ for the repetition code and $k=4$ for surface code - and the parity-based algorithms are orders of magnitude faster than the moment-based algorithms.
Moreover, parity-based algorithms also appear to scale more favorably in $E$ than moment-based algorithms.
It is, therefore, relatively clear that one should favor the use of parity-based algorithms in the low-$k$ regime.
We did not investigate DEMs with large hyperedges, but we suspect that the exponential scaling of Algorithms \ref{alg:estimate-parameters-parities} and \ref{alg:structure-learn-parities} in $k$ may render moment-based algorithms competitive.

%% Subsection: DEM Estimate Variances
\subsection{DEM Estimate Variances}
\label{subsec:dem-estimate-variances}
We do not have a closed-form expression for the variance of an arbitrary DEM excitation rate.
However, we have found the binomial standard error of the associated moment, \autoref{eq:moment-std-estimator}, to offer an accurate (and easily computable) approximation, as judged by the alignment of errors normalized by this value with the standard normal distribution (see \autoref{subsec:accuracy}).

The intuition behind this result is that the moment, $\mu_{\mathbf s}$, is a first-order function of \emph{only} $\theta_{\mathbf s}$, because all other means of producing detector response $\mathbf s$ require the excitation of more than one hyperedge.
If all $\theta \ll 1$, then higher-order terms are small and the variances of $\mu_{\mathbf s}$ and $\theta_{\mathbf s}$ are approximately equal.
This method is fast: estimating the moments of all DEM hyperedges from $10^6$ shots typically requires on the order of one second.

For completeness, we report two less satisfactory approximations of variance that we investigated: one based on a heuristic propagation of uncertainty and another using the jackknife.

For the heuristic method, we begin with the empirical observation that, when hyperedge attenuation $\psi_{\mathbf s}$ is repeatedly estimated from many different $\mathbf{X}$ sampled from the same DEM, the variance of the resulting estimates is roughly proportional to the variance of the corresponding depolarization.
Specifically,
\begin{align}
  \widehat{\mathsf{Var}}(\psi_{\mathbf s}) &\approx \frac{1}{N} \left(\frac{1}{4} \right)^{|\mathbf s|} \widehat{\mathsf{Var}}(\omega_{\mathbf s}) = \frac{1}{N} \left(\frac{1}{4} \right)^{|\mathbf{s}|} \frac{1 - \widehat{\pi}_{\mathbf s}^2}{{\widehat{\pi}}_{\mathbf s}^2}.
\end{align}
This heuristic for the variance of the attenuation can be used to approximate the variance of the excitation rate by noticing that $d\theta_{\mathbf s}/d\psi_{\mathbf s} \approx 1/4$ for small attenuations.
Therefore $\widehat{\mathsf{Var}}(\theta_{\mathbf s}) \approx \frac{1}{4} \widehat{\mathsf{Var}}(\psi_{\mathbf s})$.

This approximation is related to the fact that the \emph{aggregated} attenuation is a linear function of depolarizations, and therefore the covariance matrix of the former is a linear transform of the covariance matrix of the latter.
Assuming a diagonal covariance matrix and using the variance of aggregated attenuation to approximate the variance of the attenuation may well lead to the above result.
We do not pursue this reasoning further because the moment-based approximation in \autoref{eq:moment-std-estimator} outperforms the heuristic method.
However, like the moment-based method, this heuristic method is also fast, because it only involves computing one polarization per DEM hyperedge.
If the polarizations used in estimating the excitation rates are cached, then variance estimation by the heuristic method becomes nearly instantaneous.

The final variance estimation method uses the jackknife \cite{wu1986jacknife}.
For each of $R$ replicates, we perform Algorithm \ref{alg:estimate-parameters-parities} on $\mathbf{X}$ with a single, randomly chosen shot omitted, resulting in a set of estimates for each hyperedge excitation rate: $\lbrace \hat{\theta}_{\mathbf s}[i] \rbrace_{i=1}^R$.
The variance may then be estimated as
\begin{align}
  \widehat{\mathsf{Var}}(\theta_{\mathbf s}) &= (N - 1) \mathsf{Var}(\hat{\theta}_{\mathbf s}) = \frac{N - 1}{R} \sum_{i=1}^R (\hat{\theta}_{\mathbf s}[i] - \langle \hat{\theta}_{\mathbf s} \rangle)^2.
\end{align}
In a traditional jackknife \cite{jackknife} each sample is omitted exactly once, yielding a deterministic estimate of variance with $R=N$.
However, with syndrome corpora routinely exceeding one million shots, the traditional jackknife would be prohibitively expensive.
Since it is competing with methods that are already quite fast, we choose $R=10^3$, which requires about an order of magnitude more computation than estimating the excitation rates themselves.
Even with this setting, the jackknife is both slower and less accurate than the other two methods.

\begin{figure}
  \label{fig:variance-estimation-methods}
  \centering
  \includegraphics{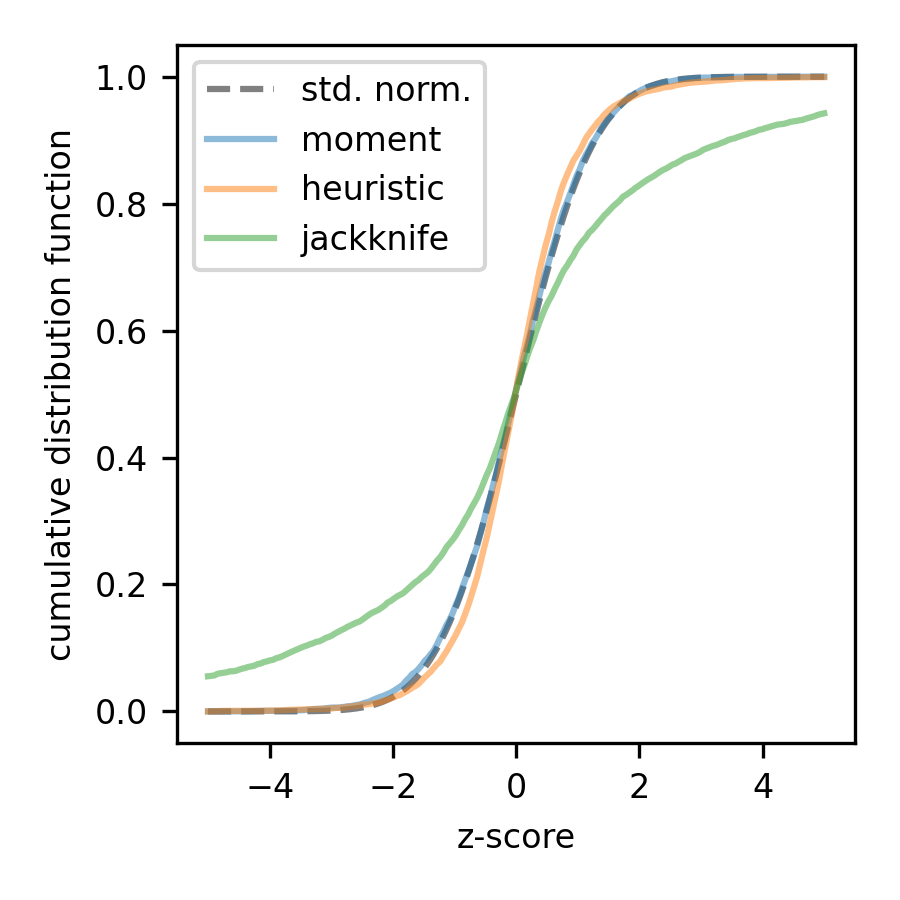}
  \caption{Comparison of methods for estimating variance of DEM hyperedge rates.
    Shown are the empirical cumulative distribution functions of residuals between estimated and true rates, normalized by the square root of the respective approximations of variance, where estimates of rates and variances are derived from $10^6$ simulated shots comprising 7 rounds of a distance-7 surface code with the SI1000 noise model.
    Better approximations adhere more closely to the standard normal distribution (gray, dashed line).}
\end{figure}

\begin{table}
  \centering
  \begin{tabular}{cccc}
    \hline
    & moment (Eq.~\eqref{eq:moment-std-estimator}) & heuristic & jackknife \\
    \hline
    variance & 1.07 & 0.95 & 11.9 \\
    skewness & -0.16 & 0.21 & 0.17 \\
    kurtosis & 0.48 & 2.92 & 21.6 \\
  \end{tabular}
  \caption{Higher moments of normalized residuals using three normalization methods.
    Residuals are computed as in \autoref{eq:normalized-moment-residuals} using $10^6$ syndromes sampled from the SI1000 DEM for 7 rounds of a distance-7 surface code.}
  \label{tab:variance-estimation-comparison}
\end{table}

Assuming estimation error is dominated by finite sample effects (see \autoref{subsec:accuracy}), then in the limit of many shots, appropriately standardized residuals should be well-approximated by a standard normal distribution.
One way to compare variance estimation methods, therefore, is to examine the higher moments of the distributions of standardized residuals that result from each method and evaluate their closeness to the standard normal distribution, which has unit variance and zero for all higher moments.
\autoref{tab:variance-estimation-comparison} shows the higher moments of normalized residuals for each of the three variance estimation methods.
Of the three, the jackknife performs the worst, consistently underestimating the actual variance of estimated rates.
Both \autoref{eq:moment-std-estimator} and the heuristic method produce normalized residuals with variance near unity and skewness near zero; however, the heuristic method has markedly higher kurtosis than \autoref{eq:moment-std-estimator}.
The standardized residuals estimated by the heuristic method are overdispersed.
When estimated variances are used in significance testing, e.g. for hyperedge discovery in Algorithms \ref{alg:structure-learn-moments} and \ref{alg:structure-learn-parities}, this overdispersity will cause the heuristic method to produce more false positives and false negatives than \autoref{eq:moment-std-estimator}.
Because it produces the most reasonable estimates and is relatively inexpensive, we use \autoref{eq:moment-std-estimator} in the remainder of this work wherever an estimate of variance is required.

%%%%%%%%%%%%%%%%%%%
% GOOGLE ANALYSIS %
%%%%%%%%%%%%%%%%%%%
\section{Analysis of Google's Willow Using DEMs}
\label{sec:analysis-of-google}

We sincerely appreciate Google releasing a large corpus of QEC data, complete with \texttt{stim} circuits, from experiments with their 72- and 105-qubit chips \cite{acharya2024quantumerrorcorrectionsurface}.
In this section, we apply the learning methods above to Google's data and demonstrate their utility in exploratory data analysis.
Some of the artifacts found by these methods are interesting in their own right, so we report them here.

%% Subsection: Data Pooling
\subsection{Data Pooling}
\label{subsec:data-processing}
In the following sections, we demonstrate effects that are only statistically significant when DEMs are estimated from several million shots.
Each experiment in the Google corpus has only $5\times 10^4$ shots, so we pool data from all experiments pertaining to the $d=7$ surface code (in each basis) on the 105-qubit chip to increase the statistical power of DEM estimation techniques.
We begin by discarding the first and last rounds of each shot prior to pooling, because detectors in these rounds are defined differently from those in the ``temporal bulk''.
We then choose a desired number of rounds per frame, $r$.
From each experiment in the $X$ and $Z$ logical basis with at least $r$ rounds in the temporal bulk, we form non-overlapping $r$-round frames from each shot; these frames then constitute our pooled syndromes.
The size of the pooled data-set depends on $r$: for $r=2$, we have $8.38 \times 10^7$ shots, whereas $r=d=7$ yields $2.35 \times 10^7$ shots.

We note that the pooling of data limits the set of possible DEM hyperedges as, by definition, DEMs cannot model events between shots.
Specifically, when modeling syndromes of $r$ rounds, the DEM may not contain a hyperedge with detectors separated by $r$ or more rounds.
Additionally, the splitting of shots into frames obscures the observation of hyperedges that span the boundary between two frames, leading to potential underestimation of excitation rates for such hyperedges.
However, when we are interested in parameter estimation of SI1000-structured DEMs, there are no hyperedges which impact more than two rounds, so these estimates will be minimally affected.

%% Subsection: Comparison of Estimated Exciation Rates
\subsection{Comparison of Estimated Excitation Rates}
\label{subsec:compare-to-rl-priors}
In addition the corresponding SI1000 DEM, each data-set in Google corpus for the $d=7$ surface code contains a DEM estimated by reinforcement learning (RL) on sub-patches of the $d=7$ surface code \cite{sivak2024optimizationdecoderpriorsaccurate}.
The SI1000 and RL-Prior DEMs provided by Google are structurally identical but differ in the excitation rates.
Starting with this structure, we compare the rates estimated by Algorithm \ref{alg:estimate-parameters-parities} to those learned by RL and those chosen \emph{a priori} in SI1000.
To enrich the comparison, we identify six hyperedge classes present in the temporal bulk of these DEMs:
\begin{itemize}
\item Point processes affecting single detectors on a space or time boundary,
\item Time-like edges between a detector and itself displaced one round,
\item Space-like edges representing data-qubit errors,
\item Space-time-like edges which go to a different detector in the next round,
\item Order 3-hyperedges and
\item Order 4-hyperedges.
\end{itemize}

\begin{figure}
  \centering \includegraphics{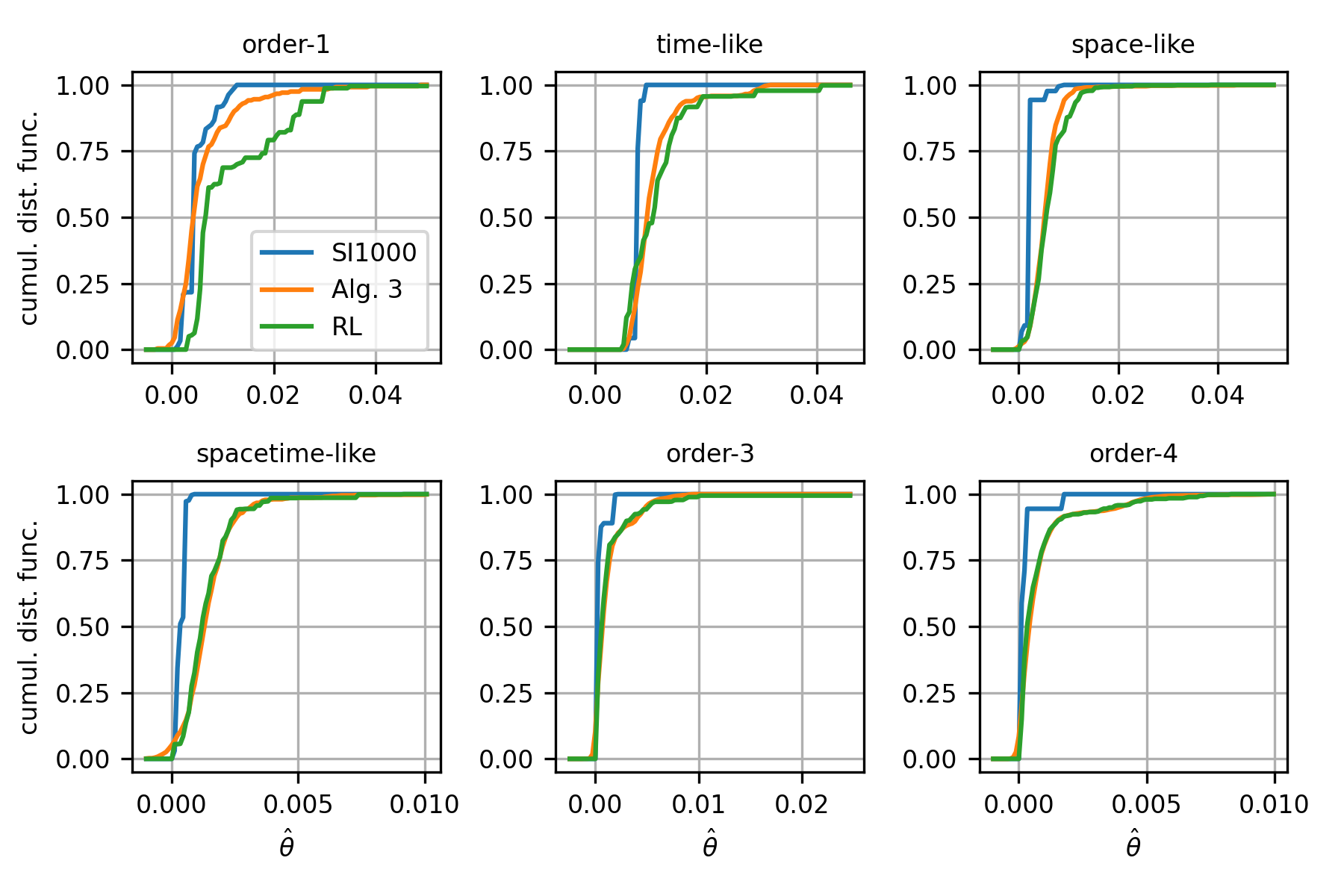}
  \caption{Cumulative distribution functions of hyperedge excitation rates.
    The six plots correspond to six classes of hyperedges discussed in the text.
    Within each plot, blue lines represent SI1000 parameters, orange lines represent parameters learned by Algorithm 3++ using syndromes from the 105-qubit chip, and green lines show parameters estimated by reinforcement learning as described in \cite{sivak2024optimizationdecoderpriorsaccurate}.}
  \label{fig:estimated-rates-vs-si1000-vs-rl}
\end{figure}

\autoref{fig:estimated-rates-vs-si1000-vs-rl} compares the three DEMs across each of the six hyperedge classes, where the data used to estimate rates for our DEM and the RL DEM come from the $d=7$, 13-round, X-basis experiment with $5 \times 10^4$ shots.
In all classes, the SI1000 rates are notably different from the other DEMs.
For space-time-like, order-3, and order-4 hyperedges, the DEMs from RL and Algorithm \ref{alg:estimate-parameters-parities} have nearly overlapping rate distributions.
For time-like and space-like hyperedges, the two DEMs are slightly different, with the RL DEM allocating slightly more mass to higher rates.
However, the largest difference appears in the order-1 hyperedges, where the rates of single-detector excitations are significantly higher in the RL DEM than when estimated by Algorithm \ref{alg:estimate-parameters-parities}.

We posit that these point-like hyperedges are most impacted because the implicit vertex in the decoding graph tied to the logical observable is connected to these detectors.
Because the RL DEM was trained to optimize logical performance, the rates of these hyperedges have an out-sized role.
We demonstrate below that the RL DEM is a better prior for downstream decoders but, at a cost of physically fidelity compared to Algorithm \ref{alg:estimate-parameters-parities}.

We quantify the effect of DEM priors on decoding performance by decoding a subset of Google's data using two decoders: PyMatching \cite{Higgot_2025} and \texttt{stim}'s BP-OSD implementation \cite{gidney2021stim,roffe_decoding_2020}, and three priors: Google's RL prior, the SI1000 prior, and our estimated prior.
At this time, we cannot use a DEM whose structure has been learned from syndromes as a prior for a decoder, because we lack an efficient means of associating a learned hyperedge with a change in the logical observable which the decoder requires.
Bridging this gap is a subject for future work.

Google's SI1000 prior is a simple, parameterized prior tailored to superconducting qubits, but not fitted specifically to a piece of hardware.
It does not differ on a qubit-to-qubit basis.
In contrast, Google's RL prior was trained as in \cite{sivak2024optimizationdecoderpriorsaccurate} and fine tuned on the provided 13-round ``calibration data-set'' taken directly from Google's 105-qubit chip.

Upon inspection of the RL hyperedge rates, it is evident that the RL priors are time-invariant.
If hyperedge $\mathbf a$ is equivalent to hyperedge $\mathbf b$ shifted by an integer number of rounds, and if neither $\mathbf a$ nor $\mathbf b$ are in the first or last round, then their rates equal.
This property presumably enabled the construction of new DEMs for different $r$-round data-sets by tiling representative hyperedges and their rates out to the required number of rounds.

For a fair comparison, we enforced time translational invariance in our DEMs via the following procedure:
\begin{enumerate}
\item For each of $X, Z$, we use Algorithm \ref{alg:estimate-parameters-parities} to estimate a DEM from the 13-round ``calibration data-set'' in the chosen basis.
\item Due to finite-sample effects, Algorithm \ref{alg:estimate-parameters-parities} can output negative rates, which are non-physical.
  For hyperedges with negative estimated rates, we replace the estimated rate with the estimated standard deviation of the corresponding moment (from \autoref{eq:moment-std-estimator}), a statistically insignificant positive value.
\item We then enforce time-invariance by averaging the resulting rates over hyperedges that are equivalent modulo time translation and do not belong to the first or last rounds.
  We call this Algorithm \ref{alg:estimate-parameters-parities}++.
\item Finally, to construct a DEM for an $r$-round experiment, we begin with the first and last rounds learned from the $r = 13$ training data and tile the time-invariant hyperedges and rates from the previous step to populate the remaining $r - 2$ time-bulk rounds.
\end{enumerate}

\begin{table}[htbp]
    \centering
    \begin{tabular}{|c|c|c|}
        \hline
        DEM & PyMatching & BP-OSD \\
        \hline
        SI1000 & $3420 \pm 81$ & $2742 \pm 73$ \\
        Alg.~\ref{alg:estimate-parameters-parities}++ & $3280 \pm 80$ & $2284 \pm 67$ \\
        RL     & $3302 \pm 80$ & $2142 \pm 65$ \\
        \hline
    \end{tabular}
    \caption{Logical error rates determined from decoding Google's $d = 7, r = 10, X$-memory surface code data-set using PyMatching and BP-OSD. Error rates are given in units of errors-per-million-rounds.}
    \label{tab:dem-decode-results}
\end{table}

In \autoref{tab:dem-decode-results}, we show decoding results for the distance-7, 10-round, $X$ basis surface code data-set.
Using PyMatching, all three priors perform comparably.
Using BP-OSD, both RL and Algorithm \ref{alg:estimate-parameters-parities} significantly outperform the SI1000 prior.
However, RL offers slightly improved performance compared to Algorithm \ref{alg:estimate-parameters-parities}.
Such a result is expected, considering the RL prior was trained to optimize logical fidelity.

%% Subsection: Divergences of DEMs from Syndromes
\subsection{Divergences of DEMs from Syndromes}
\label{subsec:kl-divergence}
The Kullback-Leibler (KL) divergence \cite{kullback-leibler-divergence} is a standard measure of disagreement between two probability distributions.
Here, we adapt it to measuring divergence between syndromes and a DEM.
In this scenario, the KL divergence of a DEM, $\mathcal{D}$, from a set of syndromes, $\mathbf{X}$, can be interpreted as the expected cost (in natural log-units) of encoding $\mathbf{X}$ using a code optimized for $\mathcal{D}$.
The KL divergence is given by
\begin{align}
  D_{\mathsf{KL}}(\mathbf{X} || \mathcal{D}) = H(\mathbf{X}, \mathcal{D}) - H(\mathbf{X})
\end{align}
where the first term is the cross-entropy, approximated as
\begin{align}
  H(\mathbf{X}, \mathcal{D}) \approx \frac{-1}{N} \sum_{\mathbf x \in \mathbf{X}} \ln \mathcal{L}(\mathbf x | \mathcal{D})
\end{align}
and the second term is the entropy of the syndromes, which does not depend on the DEM and can be estimated using sample frequencies of individual syndromes in $\mathbf{X}$.
The cross-entropy term requires the likelihood of a syndrome given a DEM, $\mathcal{L}(\mathbf x | \mathcal{D})$, which can be calculated for sufficiently small $n$ using \autoref{eq:probability-from-excitation-rates}.

As with logical performance, we desired to compare the SI1000 DEM, Google's RL DEM, and our DEMs estimated using Algorithms \ref{alg:estimate-parameters-parities} and \ref{alg:structure-learn-parities}, according to how well they agree with hardware syndromes.
However, due to the exponential complexity of evaluating $\mathcal L$, we could not use $d=7$ surface code syndromes and had to restrict our comparison to $r=2,d=3$ surface code syndromes, which were tractable with $n=16$.

We constructed two pairs of training and evaluation data-sets.
For the first training set, we took rounds 5 and 6 from each syndrome in the 13-round experiment with $d=3$ in the $X$ basis.
We did the same for the 10-round experiment in the $X$ basis to form the first evaluation data-set.
This first pair of data-sets each contain $5 \times 10^4$ shots with $r=2$ rounds per shot.
Next, we used the pooling procedure described in Section~\ref{subsec:data-processing} with $r=2$, resulting in $8.38 \times 10^7$ shots of combined $X$ and $Z$ basis syndromes.
We then iterated through the pooled data in chunks of 64 shots (an expedient dictated by the packing of detector data into 64-bit integers), placing even-numbered chunks into the pooled training data-set and odd-numbered chunks into the pooled evaluation data-set, leaving each with $4.19 \times 10^7$ shots.
These constitute the second pair of training and evaluation data-sets.

While the evaluation data-sets were used to assess all DEMs, the training data-sets were used as inputs only to Algorithms \ref{alg:estimate-parameters-parities} and \ref{alg:structure-learn-parities}, because the SI1000 DEM did not require training and the RL DEM was already trained.
Furthermore, the SI1000 and RL DEMs could not be used as-is but had to be adapted to match the two-round syndromes using the following procedure:
\begin{enumerate}
  \item Load the appropriate DEM from the $d=3,r=13,X$-basis data-set.
  \item Discard all hyperedges without any detectors in rounds 5 or 6.
  \item For each remaining hyperedge, convert to a valid hyperedge by discarding any detectors outside rounds 5 and 6.
  \item Handle collisions between valid hyperedges by summing the attenuations for all equivalent hyperedges. This is equivalent to combining their rates using the inclusion-exclusion rule.
\end{enumerate}

The Algorithm \ref{alg:estimate-parameters-parities} DEM was estimated on a training data-set with the hyperedges of the two-round SI1000 DEM from the above procedure.
Meanwhile, the Algorithm \ref{alg:structure-learn-parities} DEM was learned directly from the training data-set.

After each DEM was adapted (for SI1000 and RL) or trained (for Algorithms \ref{alg:estimate-parameters-parities} and \ref{alg:structure-learn-parities}), the KL divergence of the evaluation data-set from the DEM was computed.
The results are shown in Table \ref{tab:cross-entropy-results}.
The error intervals given in the table correspond to the standard deviation of $\ln \mathcal{L}(\mathbf x | \mathcal{D})$ scaled by $1/\sqrt{N}$, i.e. the standard error of the cross-entropy over the evaluation data-set.

Because Algorithm \ref{alg:structure-learn-parities} may learn as many hyperedges as it finds significant in the data, we were wary of over-fitting.
Evaluating KL divergence on a test set distinct from what was used to train the model implicitly penalizes over-fitting, but we also quantify over-fitting via the relative Akaike information criterion, ($\Delta \mathsf{AIC}$), which explicitly accounts for the number of hyperedges, $E$. If the absolute AIC is
\begin{align}
  \mathsf{AIC}(\mathcal{D}) = 2 \left( E(\mathcal{D}) - \ln \mathcal{L}(\mathbf{X} | \mathcal{D}) \right),
\end{align}
then the relative AIC is
\begin{align}
  \Delta\mathsf{AIC}(\mathcal{D}) = \mathsf{AIC}(\mathcal{D}) - \min_{\mathcal{D}_i} \mathsf{AIC}(\mathcal{D}_i).
\end{align}

\begin{table}[htbp]
    \centering
    \begin{tabular}{|c|ccc|ccc|}
        \hline
        & \multicolumn{3}{c|}{$N = 5 \times 10^4$} & \multicolumn{3}{c|}{$N = 4.19 \times 10^7$} \\
        $\mathcal{D}$ & $D_{\mathsf{KL}}(\mathbf{X} || \mathcal{D})$ & $E$ & $\Delta \mathsf{AIC}$ & $D_{\mathsf{KL}}(\mathbf{X} || \mathcal{D})$ & $E$ & $\Delta \mathsf{AIC}$ \\
        \hline
        SI1000 & $0.205 \pm 0.017$ & 155 & $1.2 \times 10^4$ & $0.1527 \pm 0.0006$ & 155 & $1.3 \times 10^7$ \\
        RL     & $0.146 \pm 0.013$ & 155 & $6.8 \times 10^3$ & $0.0911 \pm 0.0005$ & 155 & $7.4 \times 10^6$ \\
        Alg. \ref{alg:estimate-parameters-parities} & $0.078 \pm 0.014$ & 155 & $0$ & $0.0076 \pm 0.0005$ & 155 & $4.4 \times 10^5$ \\
        Alg. \ref{alg:structure-learn-parities} & $0.084 \pm 0.015$ & 107 & $5.3 \times 10^2$ & $0.0023 \pm 0.0005$ & 696 & $0$ \\
        \hline
    \end{tabular}
    \caption{Goodness-of-fit statistics for various DEMs relative to data-sets comprising two-round slices of Google's $d = 3$ surface code syndromes. See text for descriptions of the data-sets, columns, and error bars.}
    \label{tab:cross-entropy-results}
\end{table}

As with logical fidelity, both RL and Algorithm \ref{alg:estimate-parameters-parities} outperform SI1000 in terms of model-hardware agreement; however, whereas the RL DEM achieved better logical fidelity, the DEM estimated with Algorithm \ref{alg:estimate-parameters-parities} shows significantly better agreement with syndromes.
This result makes intuitive sense: the RL DEM was trained with a decoder in the loop in order to maximize logical fidelity, while the Algorithm \ref{alg:estimate-parameters-parities} DEM is a function of maximum likelihood estimates of syndrome statistics (depolarizations), so it ought to perform well on likelihood-based measures of model-hardware similarity.

On the smaller pair of training and evaluation data-sets, with $5 \times 10^4$ shots each, Algorithms \ref{alg:estimate-parameters-parities} and \ref{alg:structure-learn-parities} perform equally well, to within standard error, despite Algorithm \ref{alg:structure-learn-parities} learning 30\% fewer hyperedges than were present in the other DEMs (107 vs. 155).
Presumably, the rates of the omitted hyperedges were statistically unimportant to the KL divergence.

However, for the larger pair of training and evaluation data-sets, with $4.19 \times 10^7$ shots each, Algorithm \ref{alg:structure-learn-parities} learns 4.5 times as many hyperedges as are present in the other DEMs (696 vs. 155) and achieves KL divergence 1/3 the value of the nearest competitor, Algorithm \ref{alg:estimate-parameters-parities}.
From this result, we conclude that with sufficient data, the structure learned by Algorithm \ref{alg:structure-learn-parities} generalizes beyond the training data and accurately models unseen syndromes from the same hardware.
Examination of $\Delta \mathsf{AIC}$ confirms this interpretation, showing that the increase in likelihood more than makes up for the cost of the extra parameters adopted by Algorithm \ref{alg:structure-learn-parities}.

Unfortunately, the exponential complexity of $\mathcal{L}$ as a function of $n$ renders direct computation of the likelihood prohibitive for surface codes with $d=5$ and above, so this same method cannot be employed to quantify model-hardware agreement for larger codes.
However, we suspect that an ensemble of partial likelihoods computed over sub-patches of the code could be employed to similar effect, much as the RL DEM was learned by optimizing logical performance of an ensemble of sub-patches.

%% Subsection: Total Attenuation and System Stability
\subsection{Total Attenuation and System Stability}
\label{subsec:total-attenuation-system-stability}
In this section, we investigate the relationship between a simple observable - average syndrome Hamming weight - and the total attenuation of a DEM.
We apply this observation to tracking the overall performance of Google's 72-qubit device as a function of time.

Intuitively, as detection events are caused by errors, the average Hamming weight (detector event fraction) of a syndrome should track with the strength of the noise affecting the chip.
It turns out that this quantity is related to attenuation in an interesting way.
\begin{theorem}
  \label{theorem:hamming-total-atten}
  If $\mathbf x$ is a syndrome drawn from a DEM with non-zero attenuations $\psi_S$ for events $S$, then to first-order, the expected Hamming weight of $\mathbf x$ is approximately half the \emph{weighted total attenuation}:
  \begin{equation}
    2 \langle | \mathbf x | \rangle \approx \sum_S |S| \psi_S \label{eq:weighted-total-attenuation}
  \end{equation}
\end{theorem}
We give the proof in Appendix \ref{sec:proofs}.
Expected Hamming weight is an observable property of a collection of syndromes that does not depend on a DEM, so it is a model-free measure of the magnitude of noise.
\autoref{theorem:hamming-total-atten} therefore suggests the potential for an independent test for whether a DEM accounts for the observed magnitude of noise (but not whether it accurately captures spatial and temporal variations in the noise).
However, this is complicated by the approximation factor implicit in the Theorem.
This should be investigated in future works.

\begin{figure}
  \centering \includegraphics{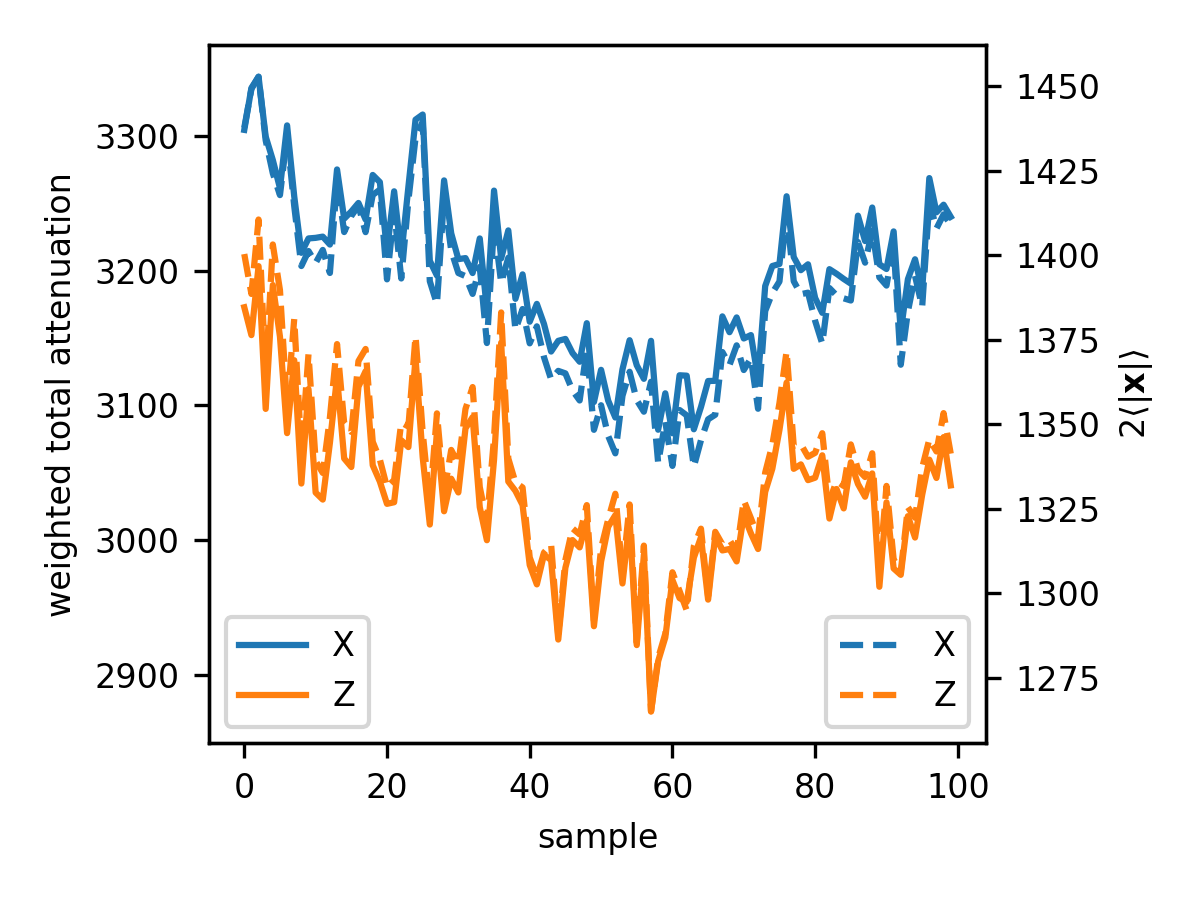}
  \caption{Weighted total attenuation (left vertical axis, solid lines) and expected syndrome weight (right vertical axis, dashed lines) vs. sample index for logical X (blue) and Z (orange) memory experiments using a distance-29 repetition code on the 72-qubit chip.
    Each plotted point is estimated from the $10^5$ syndromes in the corresponding sample.
    Weighted total attenuation is computed from DEMs estimated using Algorithm \ref{alg:estimate-parameters-parities} with the SI1000 hyperedges as input.}
  \label{fig:total-attenuation-traces}
\end{figure}

As a demonstration, we examined the data published by Google for the distance-29 repetition code executed on the 72-qubit chip.
This data is organized hierarchically as follows: the data-set comprises 100 samples in each of the logical X and logical Z bases, each sample comprises $10^5$ shots, and each shot comprises $10^3$ rounds of syndrome extraction book-ended by initialization and terminal measurement.
Using the hyperedges defined by the SI1000 DEM for a $10^3$-round shot, we chose to estimate excitation rates for each hyperedge in each $10^5$-shot sample, resulting in 200 fitted DEMs (100 in the X basis and 100 in the Z basis).
Meanwhile, we also computed the average Hamming weight per shot in each $10^5$-shot sample.
The plots of weighted total attenuation and average Hamming weight are overlaid in \autoref{fig:total-attenuation-traces}.
Note that different axes are used for the two quantities: they differ by a constant factor (beyond the expected factor of two), probably because the higher-order terms ignored by \autoref{theorem:hamming-total-atten} are non-negligible but consistent between samples.

In any case, the relative changes in the magnitude of noise across samples are evident in both metrics.
Assuming that samples are chronologically indexed, we observe that the total noise drops approximately 7\% over the first three-fifths of the samples before climbing again in the remainder.
Given that each round of syndrome extraction requires 1.1 $\mu s$, the duration of each sample is at least 110 s, and the 200 samples therefore span a minimum of 6 hours, although the actual duration would have to account for operations performed between shots and samples (such as resetting of control equipment or re-calibration).
Based on the evident overlay between X and Z curves, samples for the two bases were probably collected in an alternating fashion, i.e. one X sample followed by one Z sample or vice verse, as opposed to all X samples being collected contiguously.
The rebound in total attenuation over at least 6 hours invites speculation about diurnal error processes and urges further investigation of time-dependence of noise.

\begin{figure}
  \centering \includegraphics{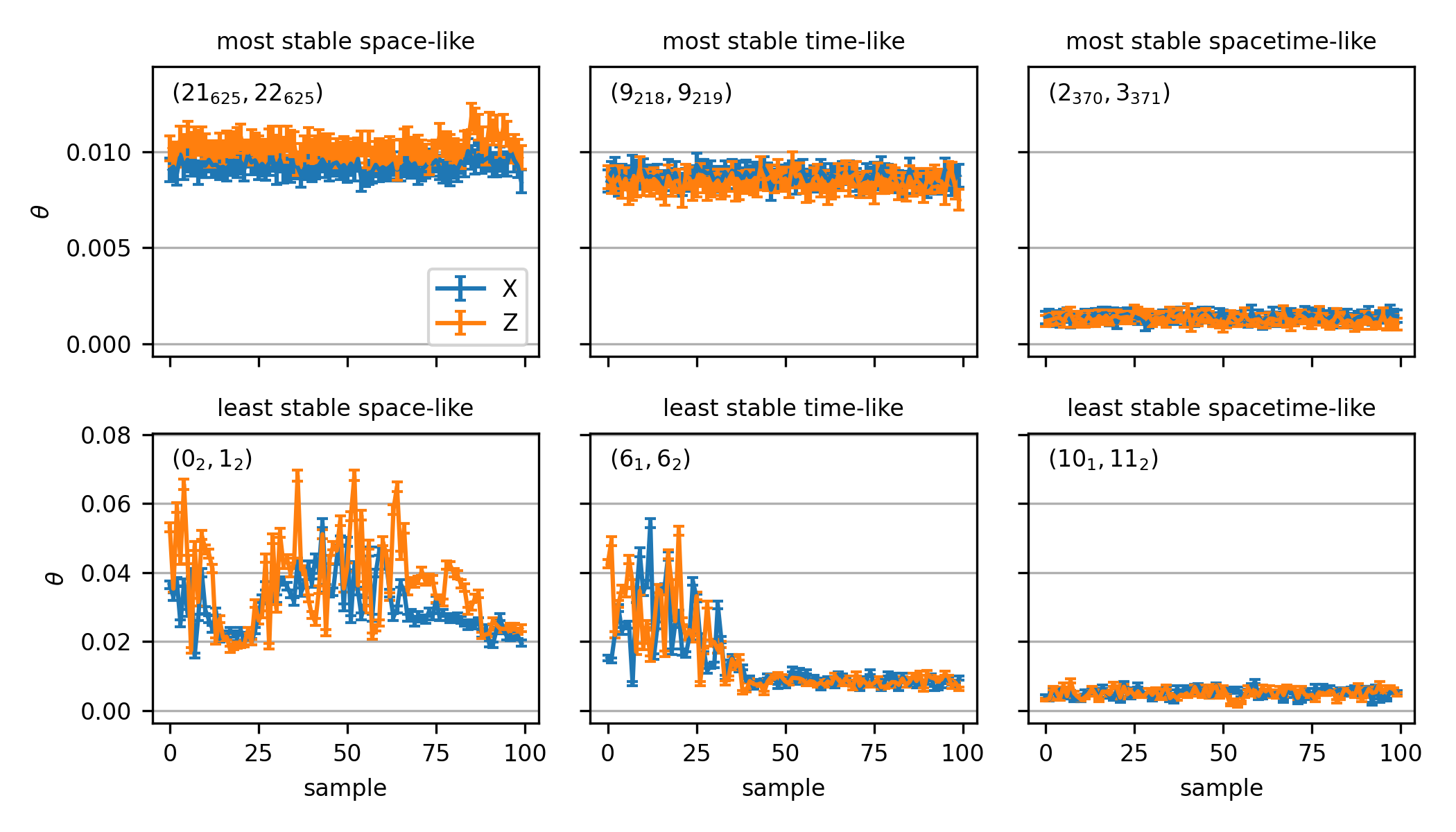}
  \caption{Rate traces for most stable (lowest variance, top row) and least stable (highest variance, bottom row) DEM hyperedges in three categories: space-like (left column), time-like (middle column), and space-time-like (right column).
    Each point represents the rate, $\theta$, of the given DEM hyperedge in the corresponding sample of the logical X (blue) or logical Z (orange) memory experiment using a distance-29 repetition code on the 72-qubit chip.
    Each plot is labeled with its DEM hyperedge as a pair of detectors, each comprising a spatial ancilla index subscripted by a round index.}
  \label{fig:stability-example-traces}
\end{figure}

We have shown that DEMs track the global magnitude of noise in syndromes, independently quantifiable in model-free statistics, but DEMs are also sensitive to local changes in the noise environment via individual hyperedge rates.
Individual DEM hyperedges show a range of stability values, as pictured in \autoref{fig:stability-example-traces}.
For the categories of space-like, time-like, and space-time-like hyperedges defined in \autoref{subsec:compare-to-rl-priors}, we found the hyperedges in each category with the lowest and highest variance across the 100 samples in the X basis, taking these as the most and least stable hyperedges, respectively, in each category.
The excitation rates of the most stable hyperedges remain effectively constant, to within estimation uncertainty, over the course of several hours.
By contrast, the least stable excitation rates may change by a factor of 3-5 for time-like and space-like hyperedges.
Interestingly, all space-time-like hyperedges appear to have high stability.

%% Subsection: Anomalies
\subsection{Anomalies}
\label{subsec:dem-anomalies}
Here, we discuss anomalies observed in Google's data-set.
First we highlight and application of the DEM formalism to identify and interpret un-modeled error correlations, resulting in a hypothesis of correlated measurement error between distant qubits.
Second, we discuss two time-dependent phenomena which violate the DEM formalism's assumptions of time-independent (or at least slowly varying) noise.
We tentatively attribute these error processes to high-energy events and possible TLS collisions.

%% Subsubsection: Correlated Measurement Error
\subsubsection{Correlated Measurement Error}
\label{subsubsec:correlated-measurement-error}
\begin{figure}
  \centering
  \includegraphics{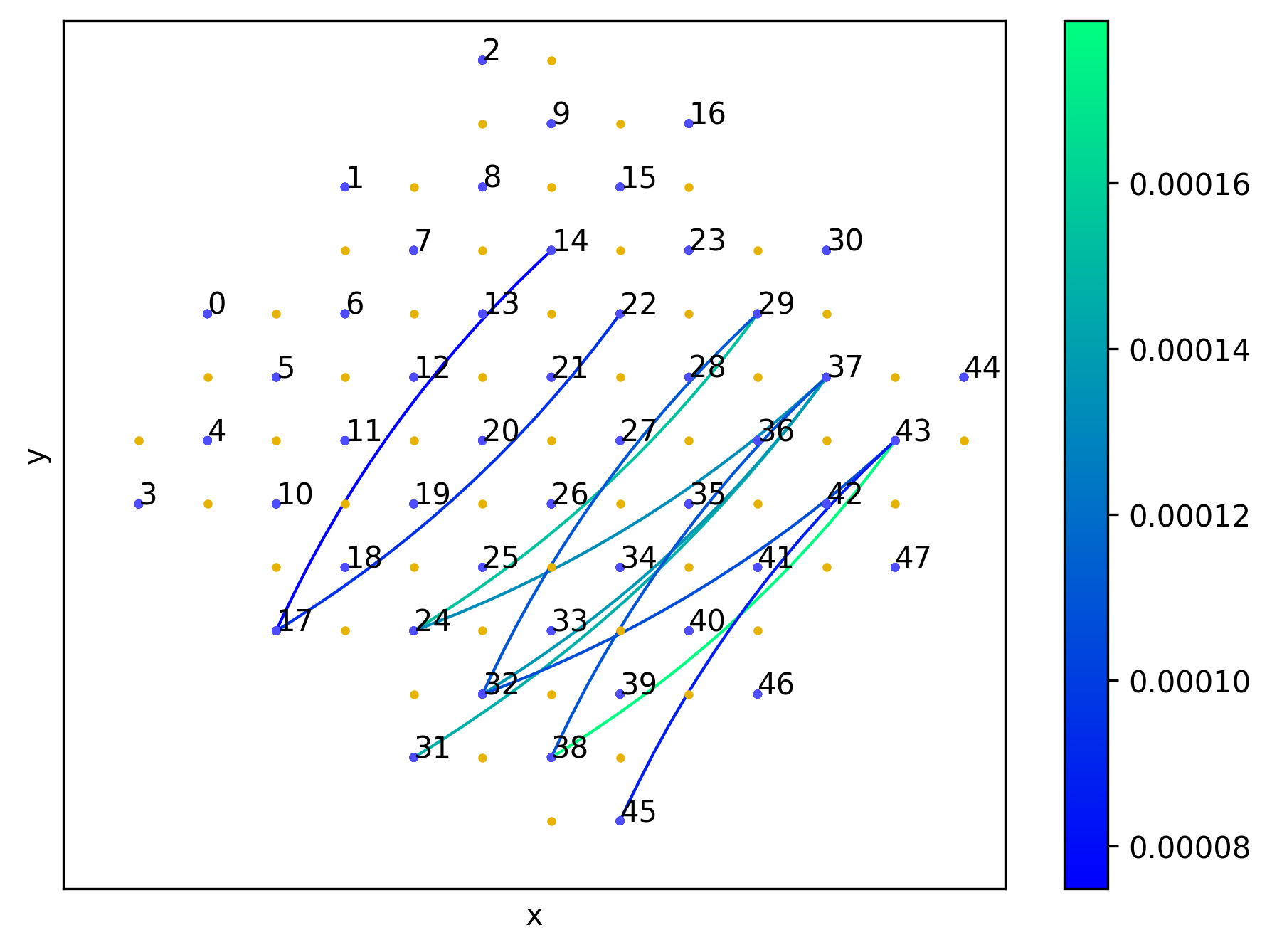}
  \caption{Statistically significant pairwise detector correlations with $\ell_1$ distance at least 8 and endpoints in the same round.
    Estimates are formed from pooled syndromes with $r=7$ for the $d=7$ surface code executed on Google's 105-qubit chip.
    Colorbar shows the excitation rate - $\theta_{\lbrace i, j \rbrace}$ - computed according to \autoref{eq:theta_ij}.}
  \label{fig:cat-scratches-pij}
\end{figure}

We now demonstrate the two-stage exploratory work-flow described in \autoref{subsec:moment-based-algorithm}.
In the first stage, we use $\theta_{\{i,j\}}$ analysis - \autoref{eq:theta_ij} and \autoref{eq:sigma_ij} - to identify statistically significant pairwise correlations.
Many such correlations are unsurprising because they arise from well-understood circuit-level noise.
The SI1000 noise model does not include any long-range correlations, for which the $\ell_1$ distance between the space-time coordinates of the endpoints exceeds 5.
Yet $\theta_{\{i,j\}}$ analysis on the distance-7, 105-qubit chip's data identifies dozens of long-range correlations.
\autoref{fig:cat-scratches-pij} shows statistically significant correlations with no temporal component (i.e. the endpoints are in the same round) and an $\ell_1$ distance of at least 8.

\begin{figure}
  \centering
  \includegraphics[width=0.8\textwidth]{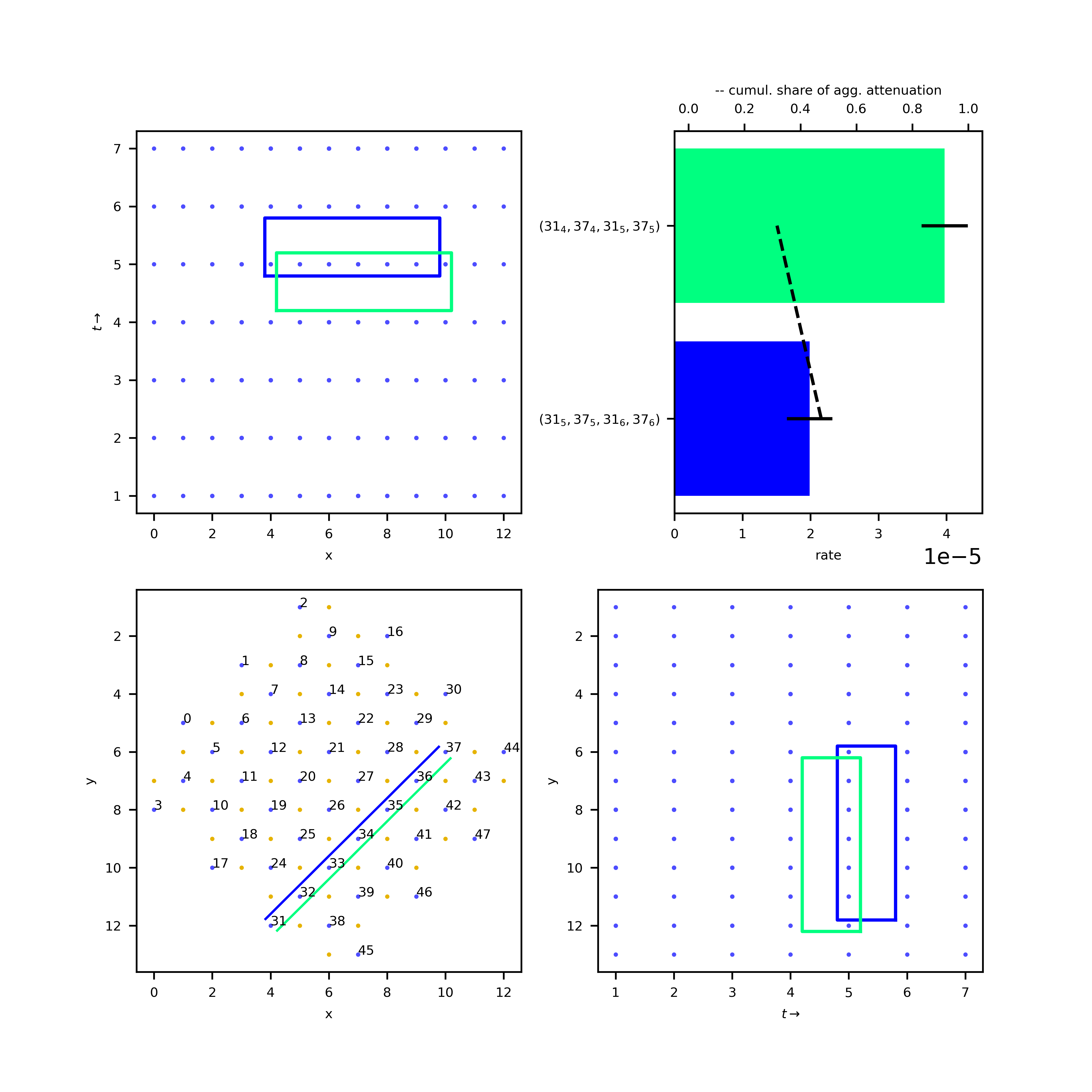}
  \caption{The output of Algorithm 4 with $\mathcal{F}=\{(31_5, 37_5)\}$, $k_{\max}=6$, and $\mathbf{X} = 2.35 \times 10^{7}$ pooled, 7-round syndromes of the distance-7 surface code on Google's 105-qubit chip.
    Lower left: statistically significant hyperedges projected onto the $x,y$ plane.
    The hyperedges are offset for clarity.
    Data qubits are shown in gold and ancillae appear in blue annotated with spatial indices.
    Upper left and lower right: analogous projections onto the $x,t$ and $t,y$ planes, respectively, where $t$ denotes the round index.
    Upper right: colored bars show the estimated rates (bottom $x$-axis) of events, whose hyperedges are plotted in the same color.
    Black lines show one estimated standard deviation above and below the estimated rate.
    The dashed line shows the cumulative fraction (top $x$-axis) of the aggregated attenuation of the seed $(31_5, 37_5)$ explained by hyperedges above and including the given hyperedge.}
  \label{fig:lineage-31-37-parities}
\end{figure}

The next step is to isolate one such correlation for further investigation.
One of the strongest pairwise correlations is between detectors with spatial indices 31 and 37.
In the pooled syndrome corpus, there are 7 such pairs, one in each round, providing further evidence for the significance of the correlation.
We avoid the first and last rounds because parent hyperedges for these pairs will be lost if they extend beyond the time boundary.
Of the 5 remaining pairs in the temporal bulk, we select the pair in round 5 because it gives the most illustrative results, labeling it $(31_5, 37_5)$.
We adopt the notation where a detector's temporal index (round number) is the subscript of its spatial index.
Using $\mathcal{F} = \{(31_5, 37_5)\}$ as a single seed and $k_{\max} = 6$, Algorithms \ref{alg:structure-learn-moments} and \ref{alg:structure-learn-parities} both find two statistically-significant, parent hyperedges: $(31_4, 37_4, 31_5, 37_5)$ and $(31_5, 37_5, 31_6,
37_6)$ with estimated rates of $4.3 \pm 0.3 \times 10^{-5}$ and $2.9 \pm 0.3 \times 10^{-5}$, respectively (error indicates one standard deviation).
As shown in \autoref{fig:lineage-31-37-parities}, these hyperedges together explain approximately 60\% of the aggregated attenuation of the seed; the other 40\% presumably belongs to other hyperedges that are either individually not statistically significant or have cardinality greater than 6.
The two significant hyperedges are consistent with correlated measurement errors on ancillae 31 and 37 in round 4, for $(31_4, 37_4, 31_5, 37_5)$, or round 5, for $(31_5, 37_5, 31_6, 37_6)$.
In the absence of other errors, a correlated measurement error would flip both detectors in one round, as each ancilla would report a different measurement than the previous round, and then both detectors would flip in the subsequent round as the ancillae reverted to the previous measurement value.

Another potential mechanism to produce a correlation between ancillae 31 and 37 is a string of alternating X and Z errors on a zig-zag path of data qubits between those ancillae.
However, such an error in round 5, for example, would \emph{only} contribute to the attenuation of $(31_5, 37_5)$ and no other parent hyperedge.
If long Pauli strings were the dominant error mechanism, then one would expect Algorithms \ref{alg:structure-learn-moments} and \ref{alg:structure-learn-parities} to find that $(31_5, 37_5)$ explains the majority of its own aggregated attenuation, without recourse to any parent hyperedges.
However, $(31_5, 37_5)$ does not appear as a statistically significant output of either algorithm.
Therefore, we conclude that long Pauli strings are not a significant contributor to the pairwise correlation between ancillae 31 and 37, the majority of which is consistent with correlated measurement error.

\begin{figure}
  \centering
  \includegraphics[width=0.8\textwidth]{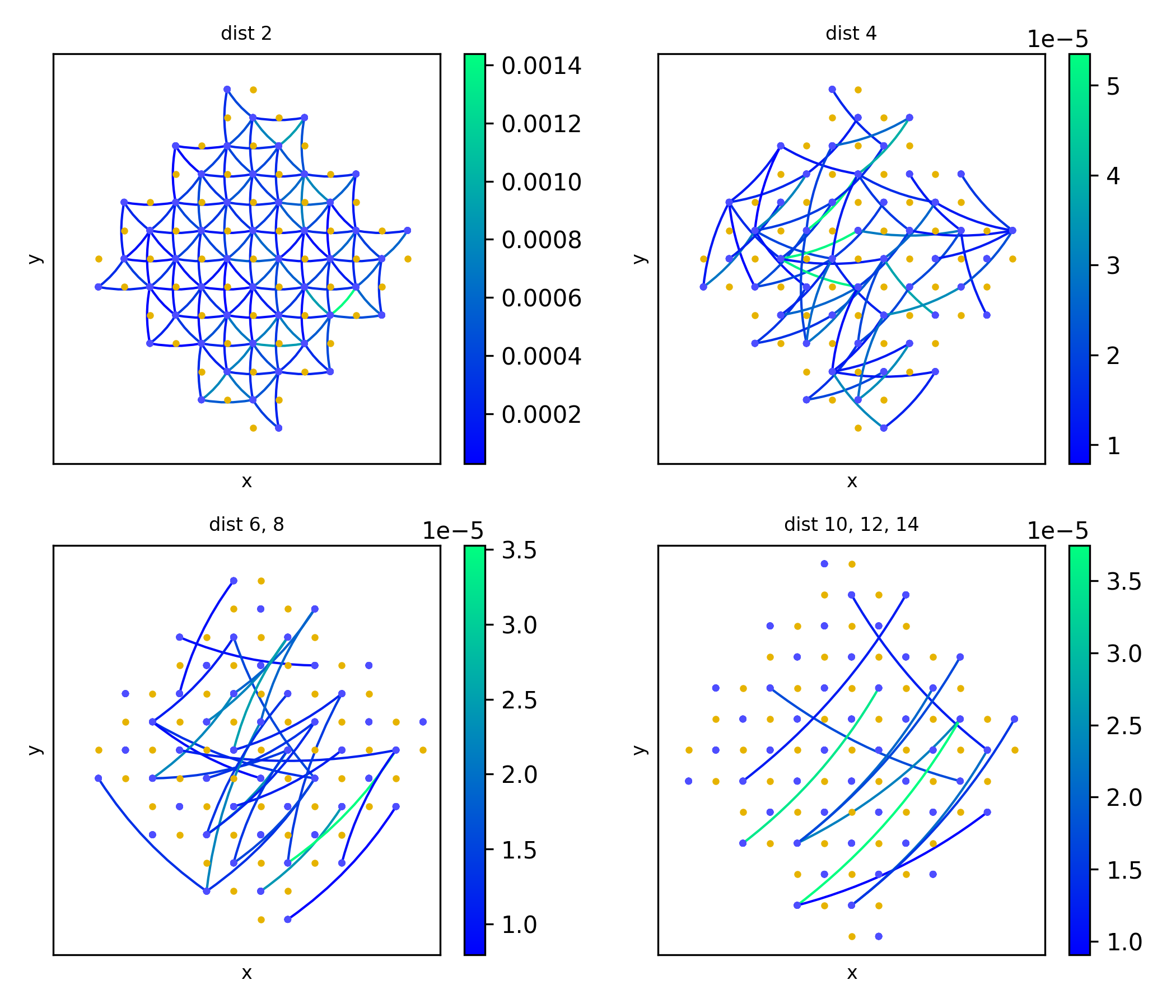}
  \caption{Statistically significant instances of the motif associated with correlated measurement error (and any other errors with the same signature), segregated by $\ell_1$ distance between the ancillae involved.
    Edge colors represent estimated rates of correlated measurement error between the ancillae joined by the edge.}
  \label{fig:correlated-measurement-error-motif}
\end{figure}

The signature of correlated measurement error between one pair of detectors led us to search for a pattern of correlated measurement errors across the chip.
In general, correlated measurement errors take the form $(a_t, b_t, a_{t+1}, b_{t+1})$ for spatial indices $a$ and $b$ and round index $t$.
We formed a DEM comprising all hyperedges that match this motif, estimated the corresponding rates using Algorithm \ref{alg:estimate-parameters-parities}, and discarded hyperedges with statistically insignificant rates.
It is important to note that the resulting hyperedge rates will include - but will not be limited to - the rates of pairwise correlated measurement errors.
For example, correlated measurement error between three ancillae will inflate the rates of motifs for each pair among the three.
There may also be unrelated errors that excite (a superset of) this motif and would also contribute to the estimated rates.
Bearing in mind these important caveats about interpretability, we can use these chip-wide motif rates to map the upper bound of correlated measurement error as a function of position on the chip.
Such a mapping is shown in \autoref{fig:correlated-measurement-error-motif}, where an edge indicates that the motif involving the ancillae joined by the edge had a statistically significant rate.
The four plots are broken out by the $\ell_1$ distance between ancillae.
The upper-left plot shows instances of the motif between nearest-neighbor ancillae, which is nearly universal, albeit likely to include localized phenomena other than correlated measurement error.
At longer ranges, the motif becomes more sparse and anisotropic, until at distances above 5 such errors are nearly all aligned along a southwest-northeast axis.
Assuming that correlated measurement error is the dominant contributor to this motif at long ranges, we strongly suspect that this pattern reflects a spatially-dependent processing of ancilla readout, perhaps in the assignment of ancillae to frequencies in a multiplexing scheme.
Verifying this suspicion would require access to details of the design of the 105-qubit chip and its control electronics.

Although this analysis does not conclusively diagnose correlated measurement error, it does nominate a strong lead for further investigation and demonstrates the utility of a multi-stage exploratory analysis starting with pairwise correlations and exploiting
various DEM estimation techniques for targeted interrogation of anomalies.
The occurrence of these and other more exotic error classes may be reflected in future error budgets estimated in the manner of \cite{acharya2024quantumerrorcorrectionsurface}.

%% Subsubsection: High-Energy Events
\subsubsection{High-Energy Events}
\label{subsubsec:high-energy-events}
In this and the following section, we highlight two classes of errors unfit for DEM analysis.
These errors are not estimable by DEM-based algorithms due to their relative rarity and long duration.
To examine the processes, we focus on the $d=29$ repetition code data collected on the gap-engineered, 72-qubit chip - by far the largest data-set in the Google corpus.
We treat the detectors as sensors rather than considering them in the context of QEC.

We define the \emph{detector event fraction} as the ratio of the number of observed detector events per the number of detectors.
The \emph{round} detector event fraction is the number of detector events out of 28 in each syndrome-extraction-cycle.
The \emph{shot} detector fraction is over all 28,056 detectors in a 1000-round shot of the repetition code memory experiment.
The \emph{qubit} detector event fraction is the average of single ancilla's detector events over a run of consecutive rounds.
\emph{Smoothed} indicates a windowed average (low-pass-filter) of the indicated quantity.

\begin{figure}[p]
  \begin{subfigure}[t]{0.5\linewidth}
    \centering\includegraphics[width=\linewidth]{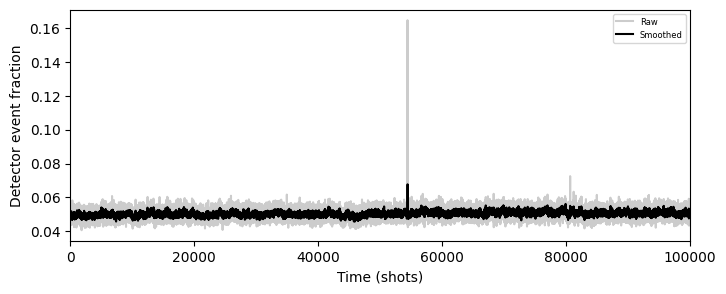}
    \caption{Shot-detection-event-fraction with a prominent candidate anomaly.
      Raw (gray) and smoothed (black) over 10 shots.}
  \end{subfigure}\hfill
  \begin{subfigure}[t]{0.5\linewidth}
    \centering\includegraphics[height=0.4\linewidth]{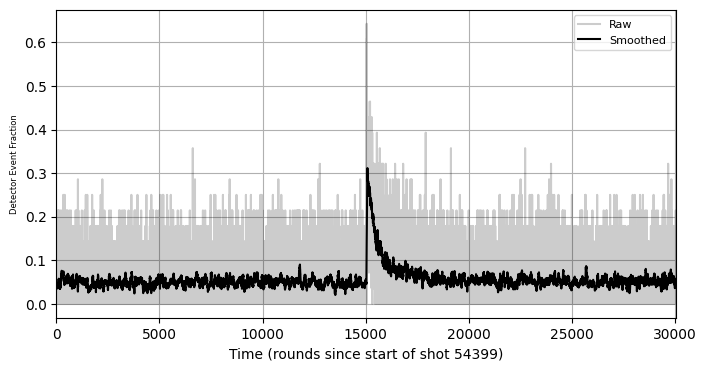}
    \caption{Zoom in on the prominent event, now plotting round-detection-event-fraction.
      Raw (gray) and smoothed (black) over 50 rounds.
      This is a likely high-energy event due to a cosmic muon.}
  \end{subfigure}
  \medskip
  \begin{subfigure}[t]{0.5\linewidth}
    \centering\includegraphics[width=\linewidth]{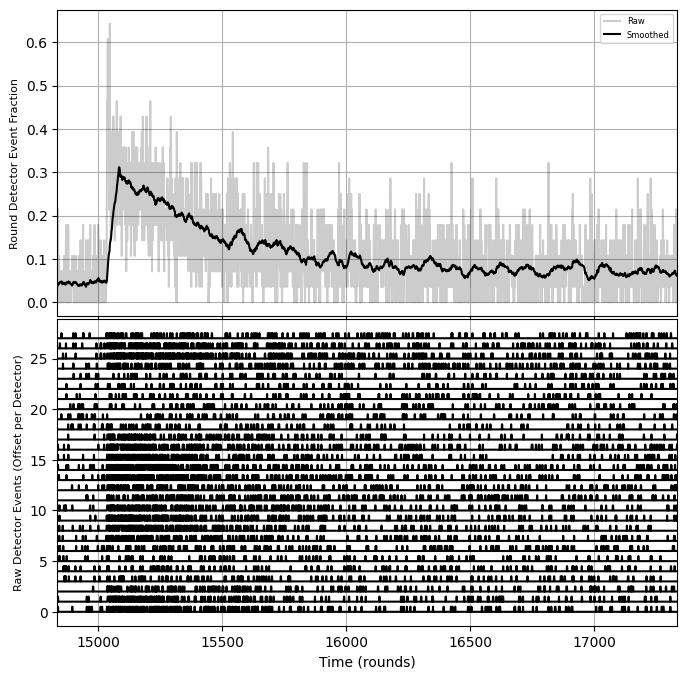}
    \caption{Closer zoom of event (above).
      Raw detector event traces (below).}
  \end{subfigure}\hfill
  \begin{subfigure}[t]{0.5\linewidth}
    \centering\includegraphics[height=\linewidth]{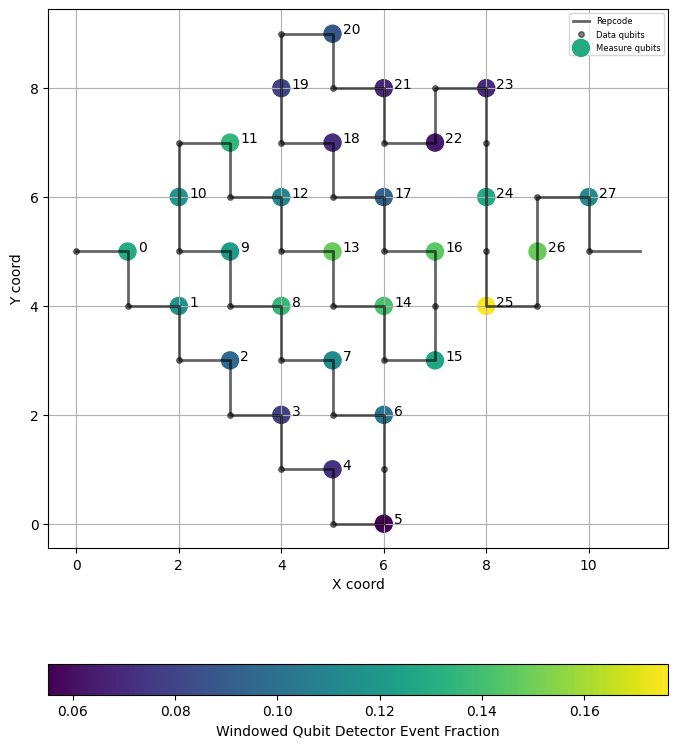}
    \caption{Qubit-detector-event-fraction from the window in (c) displayed as a heat-map on top of the repetition code.}
  \end{subfigure}
  \medskip
  \begin{subfigure}[t]{0.5\linewidth}
    \centering\includegraphics[width=\linewidth]{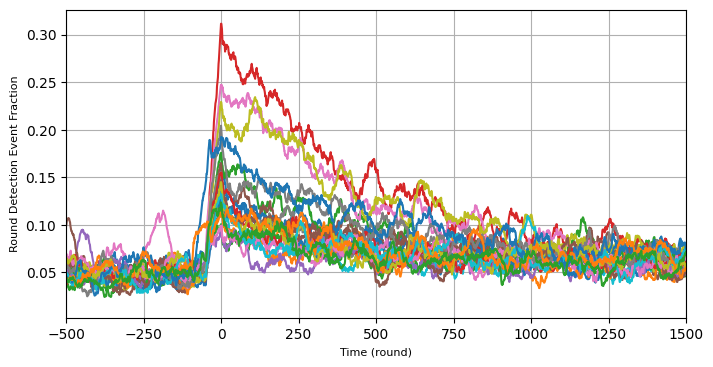}
    \caption{Overlay of all 23 detected high-energy events' smoothed round-detector-event-fraction.}
  \end{subfigure}
  \caption{Example selection of a high-energy event (a)-(d) and overlay of all selected high-energy events (e).}
  \label{fig:cosmic_example}
\end{figure}

To search for the signature of high-energy events \cite{acharya2024quantumerrorcorrectionsurface}, we locate the shot in each sample with the largest shot-detector-event-fraction in raw and smoothed (over ten shots) where peaks in both traces are within five shots.
This selection yields 30 candidates.\footnote{We also undertook a broader selection allowing more than one candidate per sample, but found no more high-energy events.
  This secondary check was useful as part of our systematic uncertainty analyses.}
We proceed to zoom in, changing time from shots to rounds, then identify the maximum round-detector-event-fraction followed by an apparent exponential falloff, examined manually.
This selection process is shown by example in \autoref{fig:cosmic_example}(a)-(c) with a spatial heat-map of qubit-detector-event-fraction in \autoref{fig:cosmic_example}(d) demonstrating this example event activity is not strictly correlated to the repetition code connectivity.

Of the 30 candidates, we identified 23 as high-energy events based on their exponential decay.
We overlay the selected 23 event windows of smoothed round-detector-event-fraction, aligned along their rising edges, in \autoref{fig:cosmic_example}(e).
Note that we discovered approximately $4\times$ more high-energy events than \cite{acharya2024quantumerrorcorrectionsurface}.
We anticipate that our threshold is looser than that publication.
Which threshold is appropriate for determining impact to QEC is a matter for future work.

%% Subsubsection: TLS-Like Events
\subsubsection{TLS-Like Events}
\label{subsubsec:tls-like-events}
\begin{figure}
  \begin{subfigure}[t]{0.48\linewidth}
    \centering\includegraphics[height=0.8\linewidth]{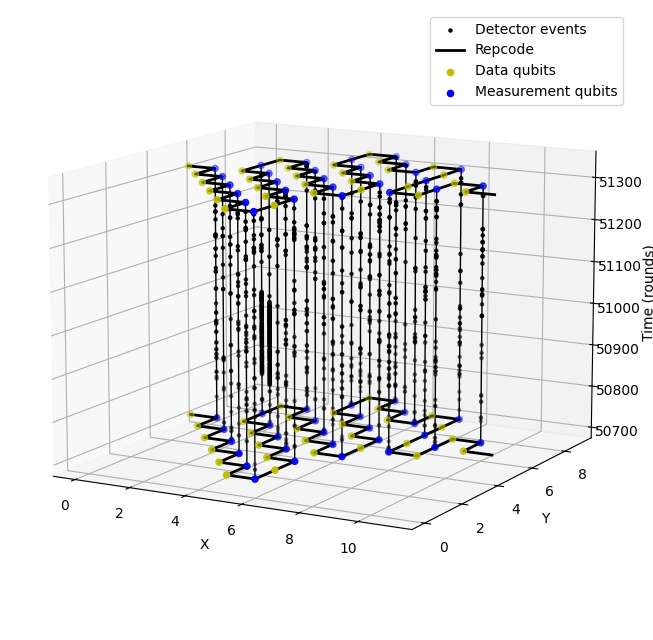}
    \caption{Example occurrence of an anomalous TLS-like event reported in the $d=29$ repetition code.
      $X, Y$ are physical coordinates on the 72-qubit chip; Z-axis is time (rounds).
      Each vertical line is a detector trace with dots representing detector events.
      The anomaly here is on detectors 9 and 10 and lasts approximately 200 $\mu$s.}
  \end{subfigure}\hfill
  \begin{subfigure}[t]{0.48\linewidth}
    \centering\includegraphics[width=\linewidth]{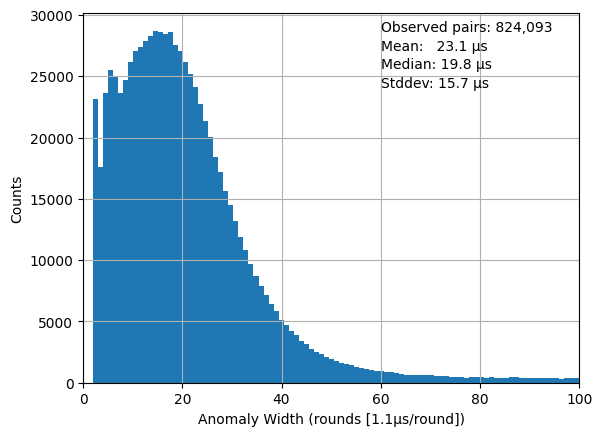}
    \caption{Distribution of approximate widths (i.e. duration) of selected TLS-like anomalies across all 200 samples.
      Average width is approximately $23 \mu$s with the maximum observed width beyond 100 syndrome extraction rounds ($>110 \mu$s).}
  \end{subfigure}
  \medskip
  \begin{subfigure}[t]{\linewidth}
    \centering\includegraphics[width=0.8\linewidth]{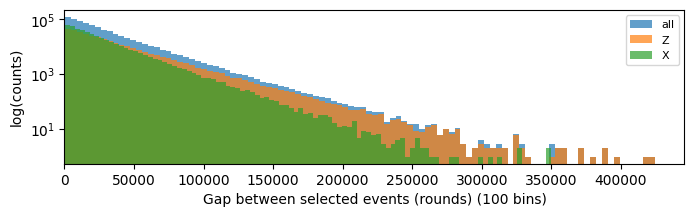}
    \caption{
      Distribution of separation of TLS-like events across all 200 samples, for all, $X$-basis, and $Z$-basis samples.
      The distributions are exponential with means: (all) $28.42 \pm 0.03$ (stat) milliseconds, ($X$) $25.13 \pm 0.04$ (stat) milliseconds, and ($Z$) $32.71 \pm 0.06$ (stat) milliseconds.}
  \end{subfigure}
  \caption{Example TLS-like error and gross timing statistics.}
  \label{fig:tls}
\end{figure}

Examining raw shot-detector-event-fraction plots such as \autoref{fig:cosmic_example}(a), after the most prominent anomalously high peaks, there are a number of further peaks well above the baseline rate of approximately 5\% to explore.
Zooming in further and moving to round-detector-event-fraction, we find a second class of anomalies more closely aligned to the repetition code connectivity, as displayed in \autoref{fig:tls}(a).
Choosing one example in round-detector-event-fraction (e.g. \texttt{sample\_05}, $Z$-basis, rounds 88,620-88,700 after the start of shot 67,400), we match-filtered all samples on similarity to this template.
This selection yielded approximately 4000 events per sample (1 per 28 ms).
We display the inter-event time distributions for all samples, then also separately, X and Z bases, in \autoref{fig:tls}(c).
We also examined the width distribution of these events by selecting the logical-and of events on pairs of neighboring detectors, counting separation between consecutive changes between detector-on to detector-off states, selecting only separations greater than two rounds, and summing the result within a window bounded by 50 rounds before and 250 rounds after the start of the identified matched-filter event (accounting for bias from any missing rounds in the selection due to momentary data-qubit-error events) to estimate the individual anomaly widths.
The resulting distribution is distinctly weakly peaked away from zero at approximately 16 rounds (18 $\mu$s) with significant numbers of events below as well as extending into a long tail of events above the peak.
Such events indicate that the underlying measurement outcomes are nearly constantly flipping over relatively long periods.
We display the full distribution in \autoref{fig:tls}(b) with relevant summary statistics about the distribution.

We hypothesize that these types of events are due to a strong coupling of the affected data qubit to some excitation in the system.
We have adopted the term TLS for their origin but acknowledge that these may be distinct from dielectric or surface defects in the physical system.
The identification of the physical origins and the impact on surface-code scalability would both be interesting lines of inquiry.

%%%%%%%%%%%%%%%%%
%% FUTURE WORK %%
%%%%%%%%%%%%%%%%%
\section{Future Work}
\label{sec:future-work}
We would like to extend these results in a few directions:
\begin{itemize}
\item We suspect that techniques from log-linear analysis, e.g. those described in \cite{10.1145/2939672.2932775}, could improve the efficiency of structure learning.
\item We seek a more generally applicable approximation of the likelihood of syndromes given a DEM.
  This would open up manifold statistical applications, such as model selection and goodness-of-fit metrics.
  The method employed above has exponential complexity in the number of detectors and is thus difficult to apply to all but the smallest codes.
  We hypothesize that the structure of the DEM could be used to decompose the syndrome into regions over which marginal likelihoods may be computed and combined.
\item We see potential applications of time-windowed DEM estimation for online operation of a logical qubit.
  For example, changes to DEM hyperedge rates over time could flag the need for re-calibration or could even provide a feedback signal to assist in online fine-tuning of qubit control.
\item We desire a method for associating learned DEM hyperedges with arbitrary-weight Pauli errors.
  Closing this gap would in principle enable a feedback loop in which estimated DEMs guide the improvement of physical noise models, which in turn lead to better QEC devices, codes, and models capable of capturing more fine-grained errors in syndromes, and so on.
\end{itemize}

Regarding this last point, our inability to associate learned DEM hyperedges with changes to the logical observable of the code is what prevents us from using the output of structure-learning algorithms as priors for decoders.
The SI1000 DEM does not face this difficulty because the SI1000 hyperedges are derived from known circuit errors whose effect on the logical observable is known.
This association between SI1000 hyperedges and the logical observable is inherited by all learned DEMs that assume the SI1000 structure, including the RL DEMs and those estimated by parameter-learning algorithms \ref{alg:estimate-parameters-moments} and \ref{alg:estimate-parameters-parities} in this work, but not by DEMs whose structure is learned from syndromes.

For learning associations between DEM hyperedges and Pauli errors, one could imagine either \emph{decoding} DEM hyperedges to Pauli errors - a step that requires information from lower-level error models to break symmetry between Pauli errors that map to the same hyperedge - or \emph{inferring} such relationships from correlations between syndromes and destructive measurements at the end of a logical experiment.
We think both approaches are worthwhile.

%%%%%%%%%%%%%%%%%
%% CONCLUSIONS %%
%%%%%%%%%%%%%%%%%
\section{Conclusions}
\label{sec:conclusions}
Both the parameters and structure of a DEM can be learned accurately from syndromes using two classes of algorithms.
Moment- and parity-based algorithms are equally accurate and achieve the maximum precision allowed by shot noise; however, they differ significantly in their run-time and scaling with the maximum cardinality of DEM hyperedges, $k$.
For DEMs with small $k$, including the repetition and surface codes employed by Google, parity-based algorithms \ref{alg:estimate-parameters-parities} and \ref{alg:structure-learn-parities} are orders of magnitude faster than moment-based algorithms \ref{alg:estimate-parameters-moments} and \ref{alg:structure-learn-moments}, with essentially no drawbacks to accuracy or precision.
Even so, we note that Algorithms \ref{alg:estimate-parameters-parities} and \ref{alg:structure-learn-parities} scale exponentially with $k$, whereas moment-based algorithms have polynomial scaling.
We speculate that there is a $k$ above which moment-based algorithms become the preferred choice.
Alternatively, parity-based Algorithm \ref{alg:estimate-from-parities-lsqr}, which does not suffer from exponential complexity in $k$, may be applicable in some scenarios.
In any case, having multiple approaches that agree is always valuable.

Arguably, the reinforcement learning employed in \cite{sivak2024optimizationdecoderpriorsaccurate} belongs to a third class of DEM estimation algorithms, albeit one that serves a different purpose than the algorithms discussed here.
On the one hand, because RL DEMs are trained with a decoder in the loop, they excel as priors for decoders used in logical memory experiments.
On the other hand, when the goal is to understand device physics, the choice of decoder becomes an unwanted degree of freedom.
As demonstrated with a $d=3$ surface code, RL DEMs do not exhibit the same fidelity to syndromes as DEMs estimated by decoder-free algorithms.

Such syndrome fidelity is important in the type of work-flow we described for discovering previously unknown classes of errors and proposing candidate mechanisms to be investigated by lower-level experiments.
The work-flow begins with the estimation of graphical DEMs comprising only pairwise detector correlations (the oldest form of DEM estimation) to identify anomalies.
The researcher then moves on to the targeted analysis of hyperedges that help differentiate between potential error classes which could explain those anomalies.
It is here that structure-learning algorithms shine.
Whereas unrestricted structure learning can result in large DEMs, which are difficult to interpret, seeded structure learning can produce digestible results that narrow the list of potential causes.

To date, most DEM analysis, including much of this paper, has assumed that hyperedge excitation rates are constant across shots (approaches differ over whether $\theta$ is allowed to vary for spatially equivalent hyperedges translated to different rounds within the shot).
However, we have also taken steps towards quantifying shot-varying effects in syndromes.
By constructing DEMs for non-overlapping windows of contiguous shots, we have shown that estimated DEM parameters are sensitive to changes in both the global and local noise environments of a QEC chip over time.
In particular, weighted total attenuation agrees with a model-independent metric showing significant changes on the scale of hours in the global noise experienced by a repetition code running on the 72-qubit chip.

Stepping outside of the DEM framework, we have also presented evidence in syndromes of spatially localized events which unfold over microseconds and are consistent with TLS interference with data-qubits.
Along with radiation events, which we find to be four times more prevalent than previously reported, these artifacts constitute a class of detector responses which is difficult to model with a DEM, because precise DEM estimation requires many thousands of shots, whereas these events span only a small number of shots.
We look forward to the continued use of DEM's as useful tools for understanding noise in QEC (and indeed other classes of quantum circuits).

%%%%%%%%%%%%%%%%
%% Bibiograpy %%
%%%%%%%%%%%%%%%%
\bibliographystyle{plain}
\bibliography{refs.bib}

%%%%%%%%%%%%%%%%
%% Appendices %%
%%%%%%%%%%%%%%%%
\appendix

\section{Proofs}
\label{sec:proofs}
We provide proofs here for results used in the main body.
We begin with Theorem \ref{theorem:GZpsi-is-Lomega} aided by two simple Lemmas.
Following that, we demonstrate the stated properties of the matrices in Table \ref{tab:matrix-properties}.
First the two Lemmas.

\begin{lemma}
  \label{lemma:discrete-independence}
  Let there be a set $S$ with subset $B$ and function $f:\mathcal{P}S\otimes\mathcal{P}S\to\mathbb{F}$ for arbitrary field $\mathbb{F}$. Then
  $$    
    \label{eq:discrete-independence}
    \sum_{A\subseteq S}f(A, B) 
    = \left[\sum_{C\subseteq S\cap B}g(C)\right]\times\left[\sum_{D\subseteq S-B}h(D)\right].
  $$
  if and only if there exist functions $g,h : \mathcal{P}S\to\mathbb{F}$ such that $f(A,B) = g(A\cap B)h(A - B)$ for all $A \subseteq S$.
\end{lemma}
\begin{proof}
Define the functions $d_B: \mathcal{P}S\to\mathcal{P}B\otimes\mathcal{P}(S-B)$ as $d_B(A) = (A\cap B, A - B)$ and $m_B: \mathcal{P}B\otimes\mathcal{P}(S-B)\to\mathcal{P}S$ as $m_B(C, D) = C \cup D$. Observe that $m_B(d_B(A)) = A$ so both functions are bijective.

Observe then:
$$
\left[\sum_{C\subseteq S\cap B}g(C)\right]\left[\sum_{D\subseteq S-B}h(D)\right] 
= \sum_{C\subseteq S\cap B}\sum_{D\subseteq S-B}g(C)h(D)
= \sum_{A \subseteq S} g(A\cap B)h(A - B)
$$
by field distributivity and the bijectivity of $d_B$ respectively.
Now if the sum and product-of-sums are equal, we move forward to see that such $g,h$ exist. Meanwhile if we assume $f(A,B) = g(A \cap B)h(A - B)$ for all $A$, we move backward and find the sum and product-of-sums are equivalent.
\end{proof}

\begin{lemma}
  \label{lemma:even-odd-subset}
  There are an equal number of even-sized and odd-sized subsets of any set. Equivalently,
  \begin{align}
    F(U) = \sum_{A \subseteq U} (-1)^{|A|} = [U = \emptyset]
  \end{align}
\end{lemma}
\begin{proof}
  Without loss of generality, we may relabel the elements of $U$ such that $U = [n]$. Let $n =
  0$. Then $U = \emptyset$ and $F(\emptyset) = (-1)^{|\emptyset|} = 1$. Now let, $U = [n + 1]$ for
  $n \ge 0$.
  \begin{align*}
    F([n + 1])
    &= \sum_{A \subseteq [n + 1]} [|A| \text{ even }]
    - \sum_{A \subseteq [n + 1]} [|A| \text{ odd }] \\
    &= \sum_{A \subseteq [n + 1]} (-1)^{|A|} \\
    &= \sum_{A \subseteq [n + 1]} (-1)^{|(A \cap [n]) \cup (A \cap \{n + 1\})|} \\
    &= \sum_{A \subseteq [n + 1]} (-1)^{|A \cap [n]|} (-1)^{[n + 1 \in A]} \\
    &= \left[\sum_{B \subseteq [n]} (-1)^{|B|}\right] \left[\sum_{C \subseteq \{n + 1\}}(-1)^{|C|}\right] \tag{Lemma \ref{lemma:discrete-independence}} \\
    &= F([n]) \times (1 - 1) \\
    &= 0
  \end{align*}
\end{proof}

Now we restate and prove Theorem \ref{theorem:GZpsi-is-Lomega}.

\begin{theorem*}
  If
  $$
  \omega_{A} = \sum_{B \subseteq [n]} (|A\cap B| \bmod 2) \psi_B,
  $$
  then for any $S \subseteq [n]$
  \begin{align}
    \label{eq:GZpsi-is-Lomega_sums_apdx}
    \frac{2^{|S|}}{-2} \sum_{S \subseteq A \subseteq [n]} \psi_A
    = \sum_{B \subseteq S} (-1)^{|B|} \omega_B.
  \end{align}
\end{theorem*}
\begin{proof}
  Starting from the RHS of Equation \ref{eq:GZpsi-is-Lomega_sums_apdx} and inserting the hypothesis:
  \begin{align*}
    \sum_{B\subseteq S}(-1)^{|B|}\omega_B
    &= \sum_{B\subseteq S}(-1)^{|B|}\left[\sum_{A\subseteq[n]}(|B\cap A|\bmod 2)\psi_A\right] = \sum_{A\subseteq[n]}\psi_A\left[\sum_{B\subseteq S}(-1)^{|B|}(|B\cap A| \bmod 2) \right] \\
    &= \sum_{A\subseteq[n]}\psi_A\left[\sum_{B\subseteq S}(-1)^{|B\cap A|+|B - A|}(|B\cap A|\bmod 2)\right] \\
    &= \sum_{A\subseteq[n]}\psi_A\left[\sum_{B\subseteq S}(|B\cap A|\bmod 2)(-1)^{|B\cap A|}(-1)^{|B-A|}\right] \\
    &= \sum_{A\subseteq[n]}\psi_A\left[\sum_{C\subseteq S\cap A}(-1)^{|C|}(|C|\bmod 2)\right]\left[\sum_{D\subseteq S-A}(-1)^{|D|}\right] \tag{Lemma \ref{lemma:discrete-independence}}\\
    &= \sum_{A\subseteq[n]} \psi_A\left[\sum_{C\subseteq S\cap A}(-1)^{|C|}(|C|\bmod 2)\right]\left[S-A=\emptyset\right]\tag{Lemma \ref{lemma:even-odd-subset}}\\
    &= \sum_{A\subseteq[n]}\psi_A\left[\sum_{C\subseteq S\cap A}(-1)^{|C|}(|C|\bmod 2)\right]\left[S \subseteq A\right] = \sum_{S\subseteq A\subseteq[n]}\psi_A\left[\sum_{C\subseteq S}(-1)^{|C|}(|C|\bmod 2)\right] \\
    &= \frac{2^{|S|}}{-2}\sum_{S\subseteq A\subseteq[n]}\psi_A.
  \end{align*}
  The last equality is due to the braced sum equaling the negative of the number of odd-sized subsets of $S$ as indicated by Lemma \ref{lemma:even-odd-subset}.
\end{proof}

We move on to the matrices of Table \ref{tab:matrix-properties}.
First we prove that the recursion given for $\mathbf{L}$ is correct.
Starting from the entry definition when $n = 0$,
\begin{align*}
  L_{00} = (-1)^{|\emptyset|} [\emptyset] = 1.
\end{align*}
Now assume that we extend our set from $[n] \to [n+1]$.
\begin{align*}
  L^{(n + 1)}_{\mathbf{a}\mathbf{b}} &= \prod_{i \in \{n + 1\}} (-1)^{b_i} [b_i \le a_i] = (-1)^{b_{n + 1}}[b_{n+1} \le a_{n+1}] \prod_{i \in [n]} (-1)^{b_i} [b_i \le a_i] \\
  &= (-1)^{b_{n + 1}}[b_{n+1} \le a_{n+1}] L^{(n)}_{\mathbf{a}^{\prime} \mathbf{b}^{\prime}} = (-1)^{a_{n+1} b_{n+1}} L^{(n)}_{\mathbf{a}^{\prime} \mathbf{b}^{\prime}}.
\end{align*}
The primed vectors above are the first $n$ bits of the respective bit-vector which we translate, using our shorthand, to integer indices.
Interestingly $\mathbf{LL} = \mathbf{I}$:
\begin{align*}
[\mathbf{LL}]_{ij} 
&= \sum_{k} L_{ik} L_{kj} \\
&= \sum_{K} \left \{ (-1)^{|K|}[K \subseteq I] \right\} \left\{ (-1)^{|J|} [J \subseteq K] \right\} = (-1)^{|J|} \sum_{K} (-1)^{|K|}[J \subseteq K \subseteq I] \\
&= (-1)^{|J|} \sum_{K} (-1)^{|J\cap K| + |K - J|}[J \subseteq K \subseteq I] = (-1)^{|J|} (-1)^{|J|} \sum_{A \subseteq I - J}  (-1)^{|A|} \\
&= \delta_{ij} \tag{Lemma \ref{lemma:even-odd-subset}}.
\end{align*}

Finally we demonstrate that $-2\mathbf{LGZ} = \mathbf{H}$ by induction. First,
\begin{align*}
\mathbf{LGZ}^{(0)} = \mathbf{L}^{(0)}\mathbf{G}^{(0)}\mathbf{Z}^{(0)} = (1) = \mathbf{H}^{(0)}.
\end{align*}
Then,
\begin{align*}
  -2\mathbf{LGZ}^{(n + 1)}
  &= -2
  \begin{bmatrix}
    \mathbf{L}^{(n)} & \mathbf{0} \\
    \mathbf{L}^{(n)} & -\mathbf{L}^{(n)}
  \end{bmatrix}
  \begin{bmatrix}
    \mathbf{G}^{(n)} & \mathbf{0} \\
    \mathbf{0} & 2 \mathbf{G}^{(n)}
  \end{bmatrix}
  \begin{bmatrix}
    \mathbf{Z}^{(n)} & \mathbf{Z}^{(n)} \\
    \mathbf{0} & \mathbf{Z}^{(n)}
  \end{bmatrix}
  = -2
  \begin{bmatrix}
    \mathbf{LGZ}^{(n)} & \mathbf{LGZ}^{(n)} \\
    \mathbf{LGZ}^{(n)} & -\mathbf{LGZ}^{(n)}
  \end{bmatrix}  \\
  &= \begin{bmatrix}
      \mathbf{H}^{(n)} & \mathbf{H}^{(n)} \\
      \mathbf{H}^{(n)} & -\mathbf{H}^{(n)} \\
    \end{bmatrix} \tag{induction hypothesis} \\
  &= \mathbf{H}^{(n+1)}.
\end{align*}

Now we restate and prove \autoref{theorem:hamming-total-atten} from the main text:
\begin{theorem*}
  If $\mathbf x$ is a syndrome drawn from a DEM with non-zero attenuations $\psi_S$ for events $S$, then to first-order, the expected Hamming weight of $\mathbf x$ is approximately half the \emph{weighted total attenuation}:
  \begin{equation*}
    2 \langle | \mathbf x | \rangle \approx \sum_S |S| \psi_S
  \end{equation*}
\end{theorem*}
\begin{proof}
  \begin{align*}
    2 \langle | \mathbf x | \rangle &= 2 \sum_{i\in [n]} \langle x_i \rangle \tag{Linearity of expectation} \\
    &= \sum_{i \in [n]} 1 - \pi_{\{i\}} \tag{Eq. \ref{eq:polarization-definition}} \\
    &= \sum_{i \in [n]} 1 - \exp(-\omega_{\{i\}}) \tag{Eq. \ref{eq:polarization-to-depolarization}} \\
    &= \sum_{i \in [n]} 1 - \left (1 - \omega_{\{i\}} + O(\omega_{\{i\}}^2)\right) \tag{Taylor expansion} \\
    &\approx \sum_{i \in [n]} \omega_{\{i\}} = \sum_{i \in [n]} \sum_S [i \in S] \psi_S \tag{Eq. \ref{eq:theorem-1-hypothesis-sums}} \\
    &= \sum_S |S| \psi_S
  \end{align*}
\end{proof}

\section{Maximum Weight of Free Excitations Required for Moment-Based Algorithms}
\label{sec:maximum-weight-in-moment-based}
\begin{figure}
  \centering
  \includegraphics{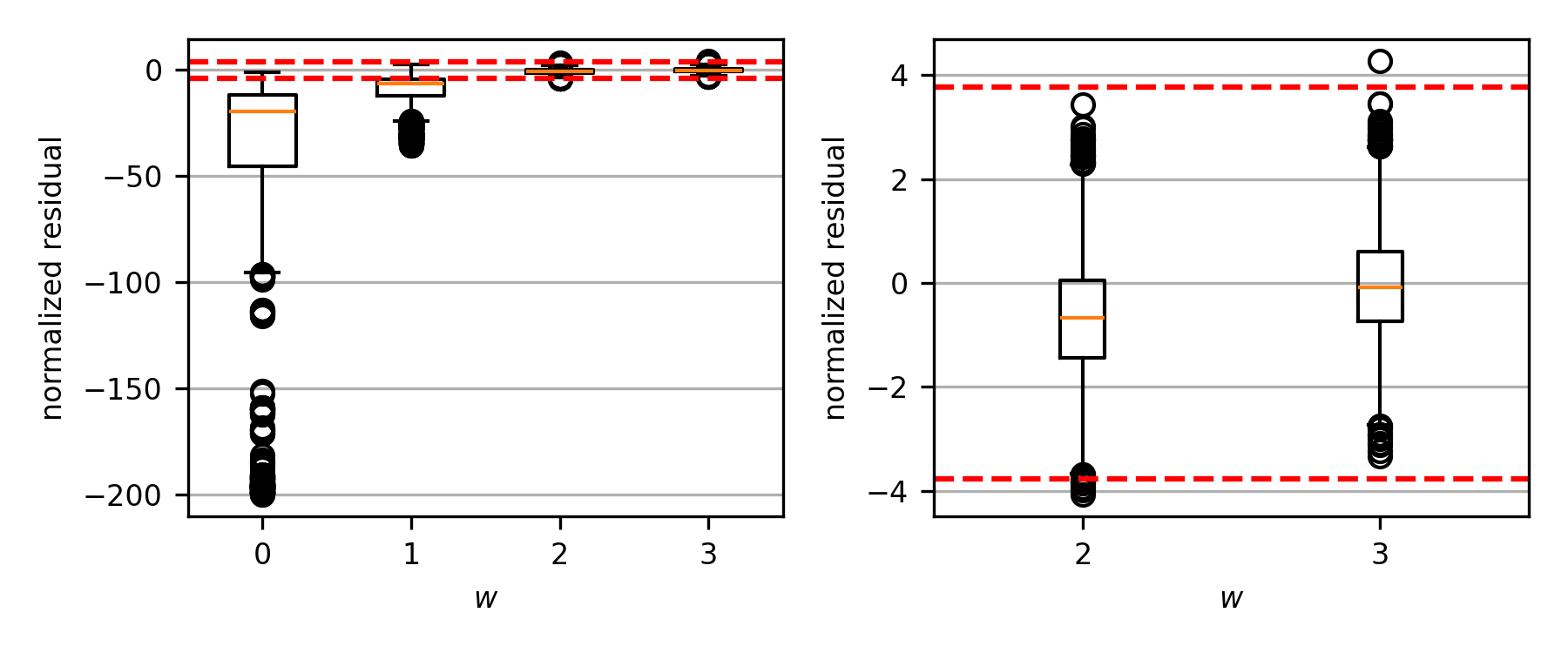}
  \caption{Distributions of normalized residuals (\autoref{eq:normalized-moment-residuals} in main text) as a function of true $\theta$ vs. maximum weight of free excitations, $w$, for $10^6$ syndromes sampled from the SI1000 DEM for $d$ rounds of the $d=7$ surface code.
  Left: $w = 0, 1, 2, 3$.
  Right: zoom in on $w = 2, 3$.
  Box-plot semantics follow \texttt{matplotlib} conventions.
  Red, dashed lines show the most extreme value expected in a sample of size $E = 5971$ drawn from a standard normal distribution.}
  \label{fig:moment-approximation-vs-free-weight}
\end{figure}

Moment-based Algorithms \ref{alg:estimate-parameters-moments} and \ref{alg:structure-learn-moments} depend crucially on $w$, the maximum weight of free excitations used to construct the excitation matrix, $\mathbf{M}^\prime$, for each hyperedge in the DEM.
Higher $w$ results in a closer approximation of the hyperedge moments via \autoref{eq:low-weight-approximate-moment}, but at exponential cost.
It is therefore critical to choose $w$ to optimize this trade-off.

One way to measure the goodness of the moment approximation is to begin with a known DEM and many syndromes sampled from it, using the true $\theta$ as input to \autoref{eq:normalized-moment-residuals} to compute normalized residuals for various $w$.
In the limit of perfect approximation, the normalized residuals will follow a standard normal distribution; therefore, any deviations can be attributed to approximation error.

\autoref{fig:moment-approximation-vs-free-weight} shows the distribution of normalized residuals vs. $w$ using $10^6$ syndromes sampled from the SI1000 DEM for $d$ rounds of the $d=7$ surface code.
This DEM has $E=5971$ hyperedges, so we expect the minimum and maximum draws from a standard normal distribution to occur at the $1/2 E$ and $1-1/2E$ quantiles, respectively, which are indicated by red, dashed lines.
For $w=0$ and $1$, the normalized residuals deviate greatly from the standard normal distribution, with medians well below zero and points far outside the expected range.
By $w=2$, the approximation is much improved: although $\tilde{\mu}(\boldsymbol{\theta})$ from \autoref{eq:low-weight-approximate-moment} is noticeably biased towards underestimating $\widehat{\mu}$, most residuals are within the expected range.
We therefore judge $w=2$ appropriate for primary tests of significance for candidate hyperedges, and we recommend this setting for the hyperedge discovery portion of Algorithm \ref{alg:structure-learn-moments}.
In parameter estimation, where precision in the estimated rates is the end goal, we recommend $w = 3$, where not only are nearly all residuals within the expected range, but the estimator is unbiased, with the median of the residuals statistically indistinguishable from zero.

\end{document}